%% file: Pupil_Masks_for_Spectrophotometry_of_Transiting_Exoplanets_en_revised_arxiv.tex
\documentclass[manuscript]{aastex}
\usepackage[dvipdfm]{}
\usepackage{amsmath}
\usepackage{here} 

\title{Pupil Masks for Spectrophotometry of Transiting Exoplanets}
\author{Satoshi Itoh, Taro Matsuo, Shohei Goda, Hiroshi Shibai, Takahiro Sumi}
\affil{Department of Earth and Space Science, Graduate School of Science, Osaka University,
1-1, Machikaneyamacho, Toyonaka, Osaka 560-0043, Japan}

\begin{abstract}
\input{PMSTE_abstract}

\end{abstract}
\begin{document}
\maketitle
\newpage
\section{Introduction}
\input{PMSTE_introduction}

\section{Theory}
\input{PMSTE_theory_header}

\subsection{Preparation}
\input{PMSTE_theory_preparation}
\subsection{Analytical expression for field-stop-loss variation}
\subsubsection{1-D apertures}
\input{PMSTE_theory_1-D}

\subsubsection{Circular isotropic apertures}
\input{PMSTE_theory_CI}
\subsection{Performance metrics}
\input{PMSTE_theory_PerMet}

\subsection{Conventional apodizing masks}
\input{PMSTE_theory_ApoMas}

\section{New Type of Masks}
\input{PMSTE_new_header}
\subsection{Mathematical expression}
\subsubsection{1-D apertures}
\input{PMSTE_new_ME_1-D}
\subsubsection{Circular isotropic apertures}
\input{PMSTE_new_ME_CI}
\subsection{\label{performance}Performance}
\subsubsection{1-D apertures}
\input{PMSTE_new_perf_1-D}
\subsubsection{Circular isotropic apertures}
\input{PMSTE_new_perf_CI}

\section{Discussion}
\input{PMSTE_discussion_header}
\subsection{Impact of optical aberrations}
\input{PMSTE_discussion_IOA_header}
\subsubsection{Impact of high-order phase aberration}
\input{PMSTE_discussion_IOA_HOA}

\subsection{Impact of mask imperfections}
\input{PMSTE_discussion_IMI_header}
\subsubsection{Transmittance deviation}
\input{PMSTE_discussion_IMI_TD}
\subsubsection{Positional deviation of the transmittance-changing point}
\input{PMSTE_discussion_IMI_PD}
\subsection{Other possible uses}
\input{PMSTE_discussion_OU}
\section{Conclusion}
\input{PMSTE_conclusion}

\vspace{25pt}
\input{PMSTE_acknowledgement}

\appendix
\renewcommand{\theequation}{A.\arabic{equation}}
\setcounter{equation}{0}
\section{Derivation of the analytical expression for field-stop-loss variation}
\subsection{1-D apertures\label{1-D_ana}}
\input{PMSTE_appendix_DAEFV_1D}
\subsection{Circular isotropic apertures\label{CI_ana}}
\input{PMSTE_appendix_DAEFV_CI}
\renewcommand{\theequation}{B.\arabic{equation}}
\setcounter{equation}{0}
\section{Mathematical formulation of block-shaped masks\label{both_MF}}
\subsection{1-D apertures\label{1-D_MF}}
\input{PMSTE_appendix_MFBSM_1D}
\subsection{Circular isotropic apertures\label{CI_MF}}
\input{PMSTE_appendix_MFBSM_CI}
\begin{table}[H]
\begin{center}

\label{eclipse}
\begin{tabular}{c c c}
\hline \hline
Stellar effective temperature & Secondary eclipse & Transmission spectroscopy \\
\hline
3200 K&8.9 ppm&12 ppm\\	
\hline
3300 K&4.7 ppm&6.5 ppm\\
 \hline
3400 K&3.0 ppm&4.3 ppm\\ 
\hline
3500 K&2.1 ppm&3.1 ppm\\ 
\hline \hline
\end{tabular}
\end{center}
\caption{Eclipse depths for transit observations of early and mid M-type stars. 1 Earth-radius is assumed as planetary radius. The effective temperature and the effective atmospheric scale height of the planet are fixed to 300 K and 20 km, respectively. Empirical polynomial equation for the relationship between stellar effective temperature and radius derived in Boyajian et al. (2012) is used. }
\label{depth}
\end{table}

\begin{figure}[H]
\begin{center}
\includegraphics[width=140mm]{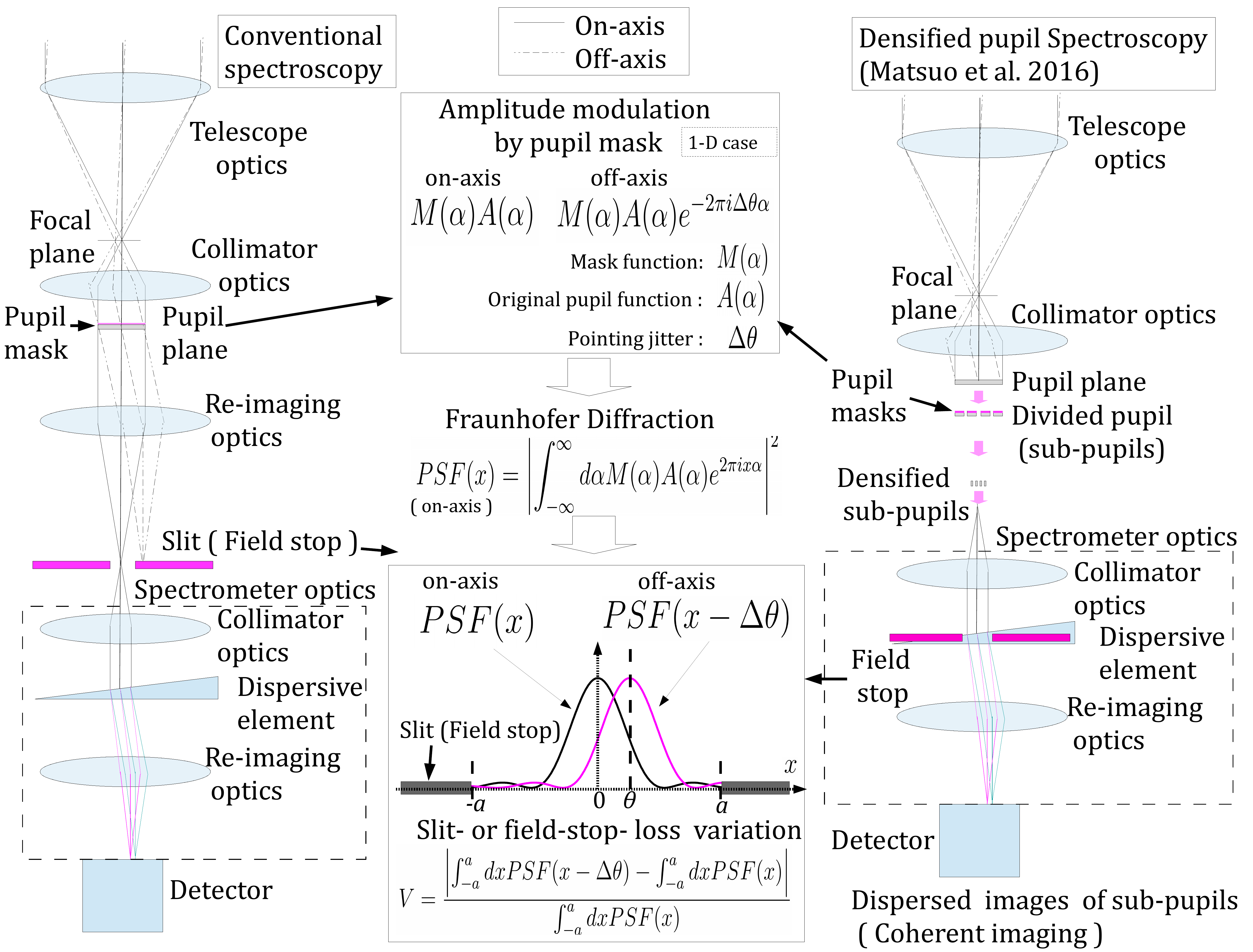}
\caption{\label{LAYOUT} Schematic of the layout of the pupil mask and field stop for conventional spectroscopy and densified pupil spectroscopy. A real pupil plane is needed for inserting a pupil mask before the field stop. By normalization, $a$ and $\Delta\theta$ are proportional to $\lambda^{-1}$ for a certain constant physical angle. This means that, for polychromatic light with a wide wavelength range, field stops with radii proportional to the wavelength of light are favorable for block-shaped masks (Section. \ref{performance}).}
\end{center}
\end{figure}

\begin{table}[H]
\begin{center}
\begin{tabular}{c l}
\hline \hline
Symbol & Meaning \\
\hline
$\lambda$ & wavelength of light \\	
\hline
$d_{max}$ & pupil radius before the use of masks\\
 \hline
$D$ & pupil diameter before the use of masks (=2$d_{max}$) \\
 \hline
$V$ & photometric variation \\
\hline
\shortstack{$x$ \\ \ } & \shortstack{image plane angular position normalized by $\lambda/D$ \\ for 1-D apertures}\\ \hline
\shortstack{$r$ \\ \ } & \shortstack{image plane angular radial coordinate normalized by $\frac{j_{1,1}}{\pi}\lambda/D\approx1.22 \lambda/D$ \\ for circular isotropic apertures} \\ \hline
\shortstack{$a$ \\ \ } & \shortstack{field-stop angular radius normalized by $\lambda/D$ for 1-D apertures\\ or $1.22 \lambda/D$ for circular isotropic apertures} \\ \hline
\shortstack{$\Delta \theta$ \\ \ } & \shortstack{pointing jitter normalized by  $\lambda/D$ for 1-D rectangular apertures\\ or $1.22 \lambda/D$ for circular isotropic apertures} \\ \hline
$\delta_{loss}$ & required photometric stability (threshold for $V$) \\ \hline
$\tau$ & energy transmittance of pupil mask \\ \hline
 \hline
\end{tabular}
\end{center}
\caption{Notation of symbols}
\label{notation}
\end{table}

\begin{table}[H]
\begin{center}
\begin{tabular}{c c}
\hline \hline
V (mag) & Number density $(\mathrm{arcsec}^{-2}\mathrm{mag}^{-1})$ \\
\hline
15.0 &$3\times 10^{-5}$\\	
\hline
17.5 &$1\times 10^{-4}$\\
 \hline
20.0 &$6\times 10^{-4}$\\ 
\hline
22.5 &$4\times 10^{-3}$\\ 
\hline
25.0 &$2\times 10^{-2}$\\
 \hline
27.5 &$6\times 10^{-2}$\\ 
\hline
30.0 &$2\times 10^{-1}$\\ 
\hline \hline
\end{tabular}
\end{center}
\caption{Relation between V-band magnitude and number density of galaxies observed in Hubble ultra deep field.  The empirical relation derived in Conselice et al. (2016) was used.}
\label{HUDF}
\end{table}

\begin{table}[H]
\begin{center}
\begin{tabular}{c c c c}
\hline \hline
V (mag) &$\mathrm{V-V_{abs}}$& $z$ & $\theta$ (arcsec) \\
\hline
15.0 &36.0&0.03&10\\	
\hline
17.5 &38.5&0.1&3\\
\hline
20.0 &41.0&0.3&2\\	
\hline
22.5 &43.5&0.6&0.7\\
\hline
25.0 &46.0&2&0.8\\	
\hline
27.5 &48.5&5&1\\
\hline
30.0 &51.0&15&2\\	
\hline 
\hline
\end{tabular}
\end{center}
\caption{The angular diameter such that the apparent magnitude of a galaxy whose size and absolute magnitude is 7 kpc and V$=-21$ ($\approx$ M31) becomes $V$. The empirical relation between $z$ and $\mathrm{V-V_{abs}}$ indicated in Figure.4 of Riess et al. 2004 was used.}
\label{M31}
\end{table}

\begin{table}[H]
\begin{center}
\begin{tabular}{c c c c}
\hline \hline
V (mag) &$\mathrm{V-V_{abs}}$& $z$ & $\theta$ (arcsec) \\
\hline
15.0 &31.0&0.003&9\\	
\hline
17.5 &33.5&0.01&3\\
\hline
20.0 &36.0&0.03&1\\	
\hline
22.5 &38.5&0.1&0.3\\
\hline
25.0 &41.0&0.3&0.2\\	
\hline
27.5 &43.5&0.6&0.07\\
\hline
30.0 &46.0&2&0.08\\	
\hline 
\hline
\end{tabular}
\end{center}
\caption{The angular diameter such that the apparent magnitude of a galaxy whose size and absolute magnitude is 0.7 kpc and -16  becomes $V$. The empirical relation between $z$ and $\mathrm{V-V_{abs}}$ indicated in Figure.4 of Riess et al. 2004 was used.}
\label{M31over10}
\end{table}

\begin{figure}[H]
\begin{center}
\includegraphics[width=80mm]{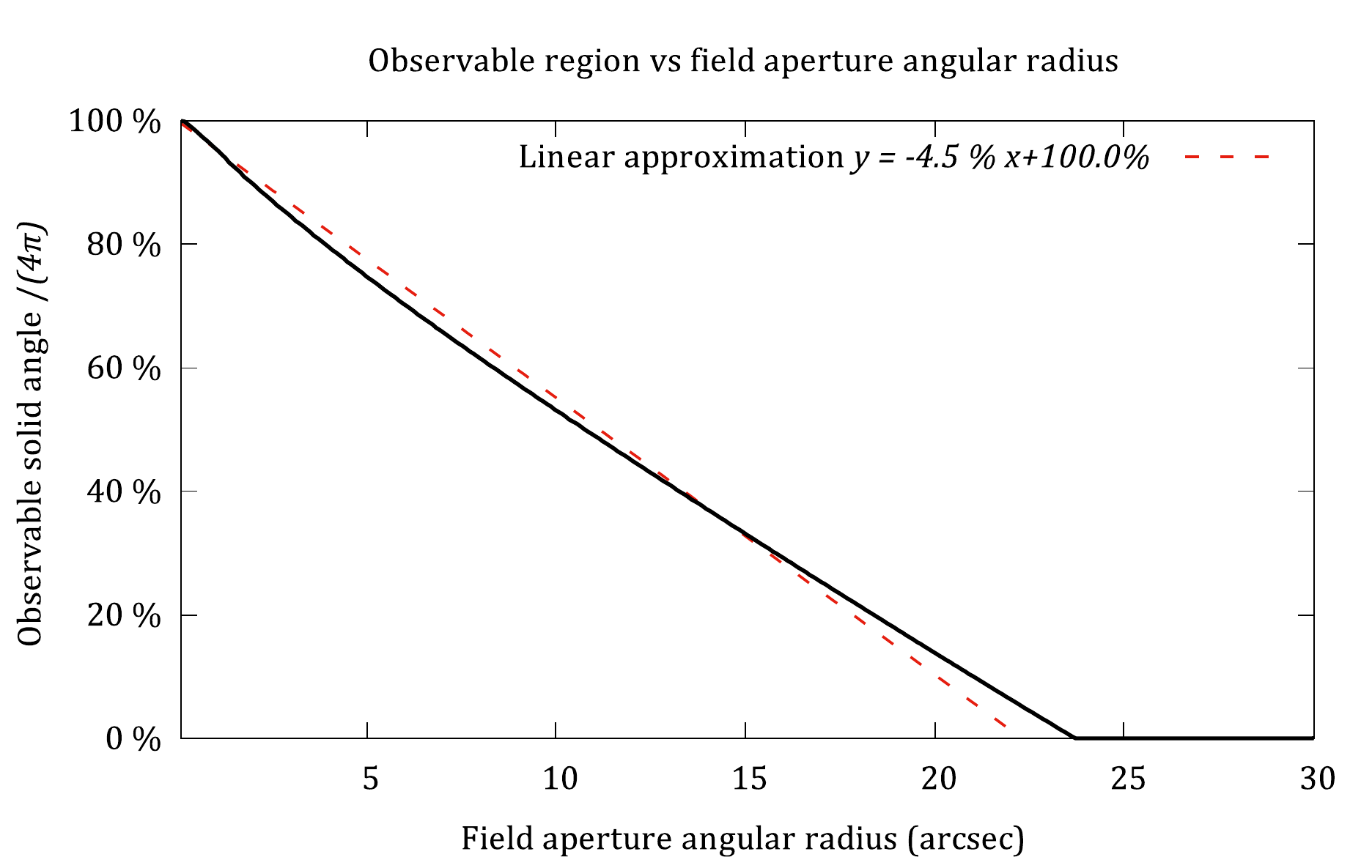}
\caption{\label{SAvsFA}Ratio of observable solid angle to $4 \pi$ as a function of field aperture angular radius (Only stars in the Milky Way are considered.)}
\end{center}
\end{figure}

\begin{table}[H]
\begin{center}
\begin{tabular}{c c c c c}
\hline \hline
Case & Function type & Parameters & \shortstack{Energy transmittance \\ (1-D)}& \shortstack{Energy transmittance \\(Circular isotropic)} \\
\hline
(A) & no masks & & 100\ \%  & 100\ \% \\	
\hline
(B) & Gaussian & $\sigma$=0.4 & 25\ \% & 8\ \% \ \\
 \hline
(C) & Gaussian & $\sigma$=0.5 & 31\ \%  & 12\ \% \ \\
 \hline
(D) & Gaussian & $\sigma$=0.6 & 37\ \%  & 18\ \% \ \\
 \hline
(E) &Hyper-Gaussian&a'=0.72 b'=0.19 n=4 & 86\ \%  & 75\ \% \\ 
\hline
(F) &Hyper-Gaussian&a'=0.82 b'=0.09 n=4 & 89\ \%  & 88\ \% \\ 
\hline \hline
\end{tabular}
\end{center}
\caption{\label{mask}List of tested apodizing masks}
\end{table}

\begin{figure}[H]
\begin{center}
\includegraphics[width=80mm]{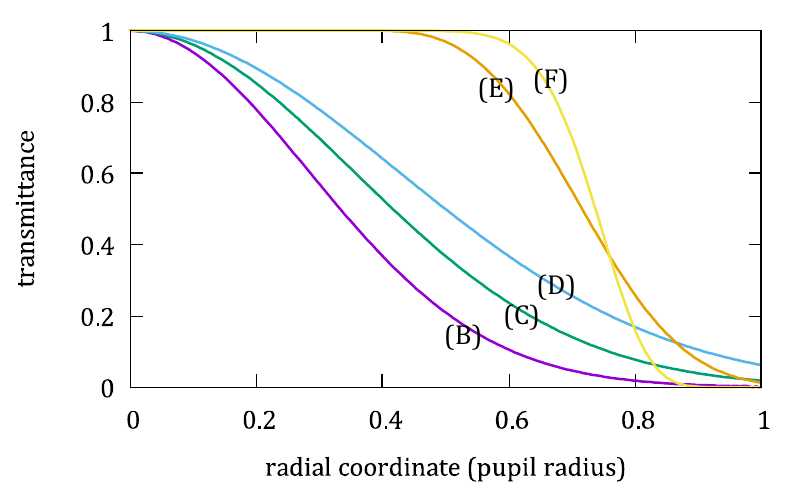}
\caption{\label{PUPIL_MASK}Amplitude transmittance of masks as a function of pupil radius}
\end{center}
\end{figure}

\begin{figure}[H]
\begin{center}
\includegraphics[width=80mm]{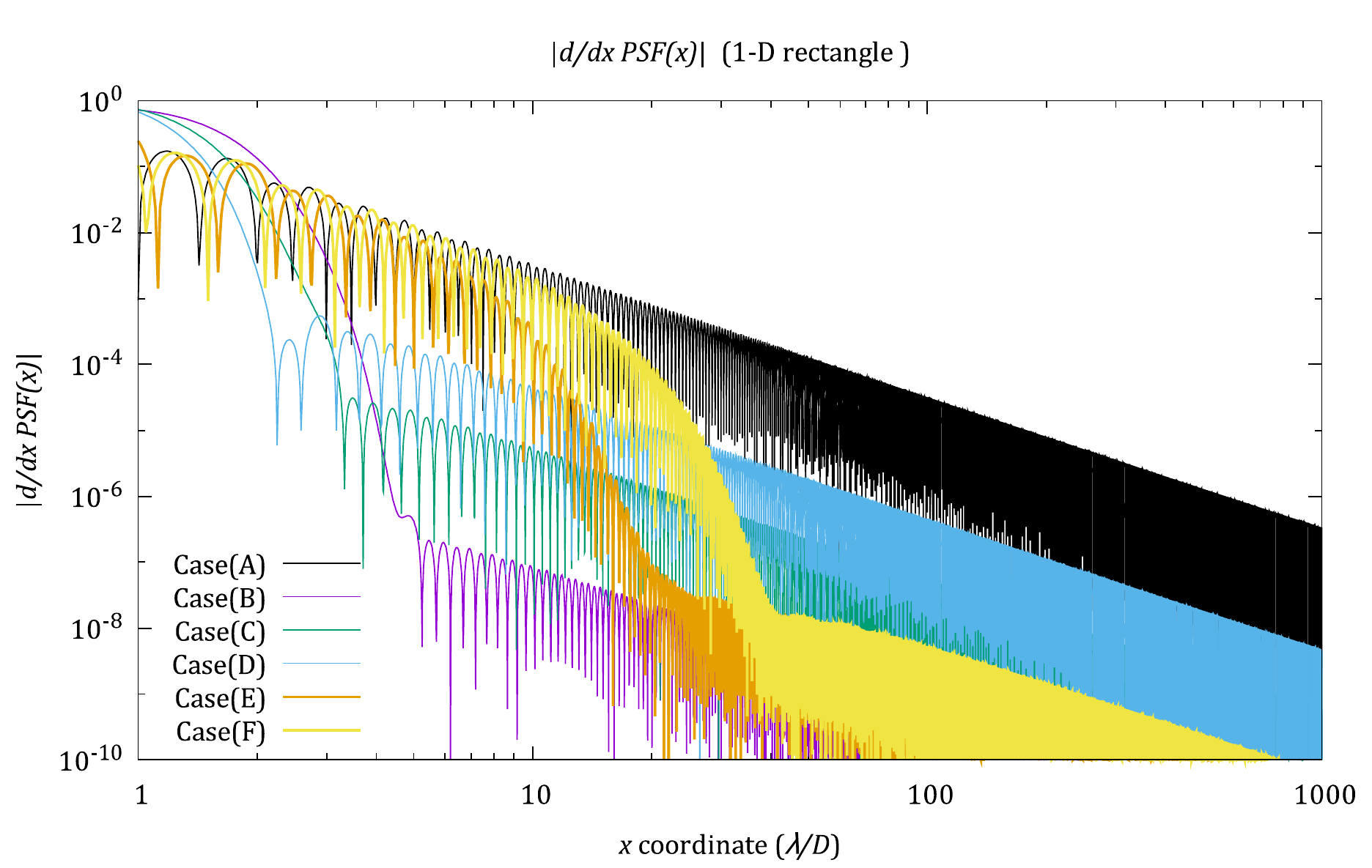}
\caption{\label{PSF_dash_1D_log} $ \left|\frac{d}{dx} \protect\scalebox{0.7}{ $\displaystyle PSF(x) \mid_{x=a}$ }\right|$ for 1-D rectangular apodizing masks}
\end{center}
\end{figure}

\begin{figure}[H]
\begin{center}
\includegraphics[width=80mm]{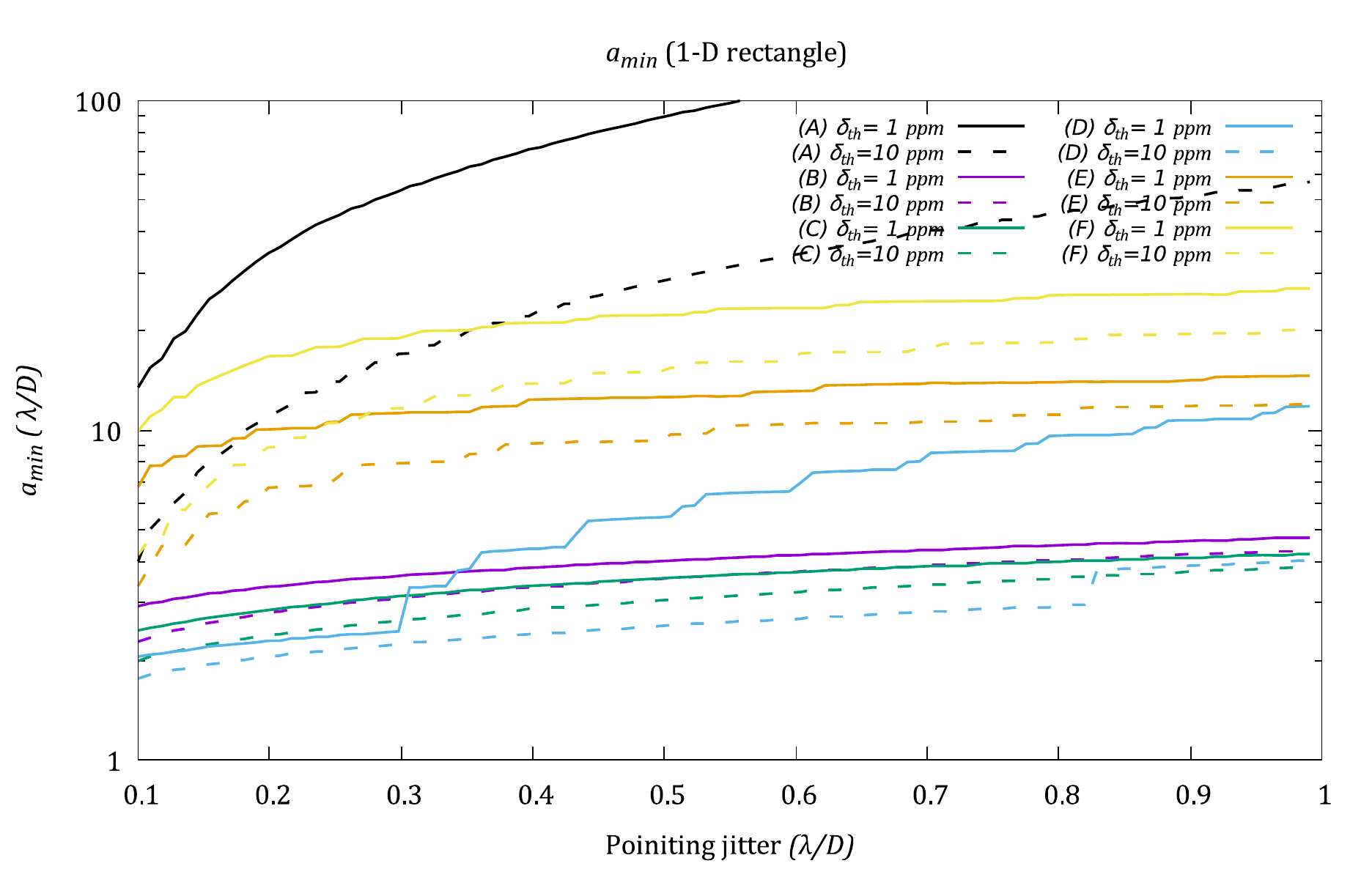}
\caption{\label{a_min_1D_log} $a_{min}$ for 1-D rectangular apodizing masks}
\end{center}
\end{figure}

\begin{figure}[H]
\begin{center}
\includegraphics[width=80mm]{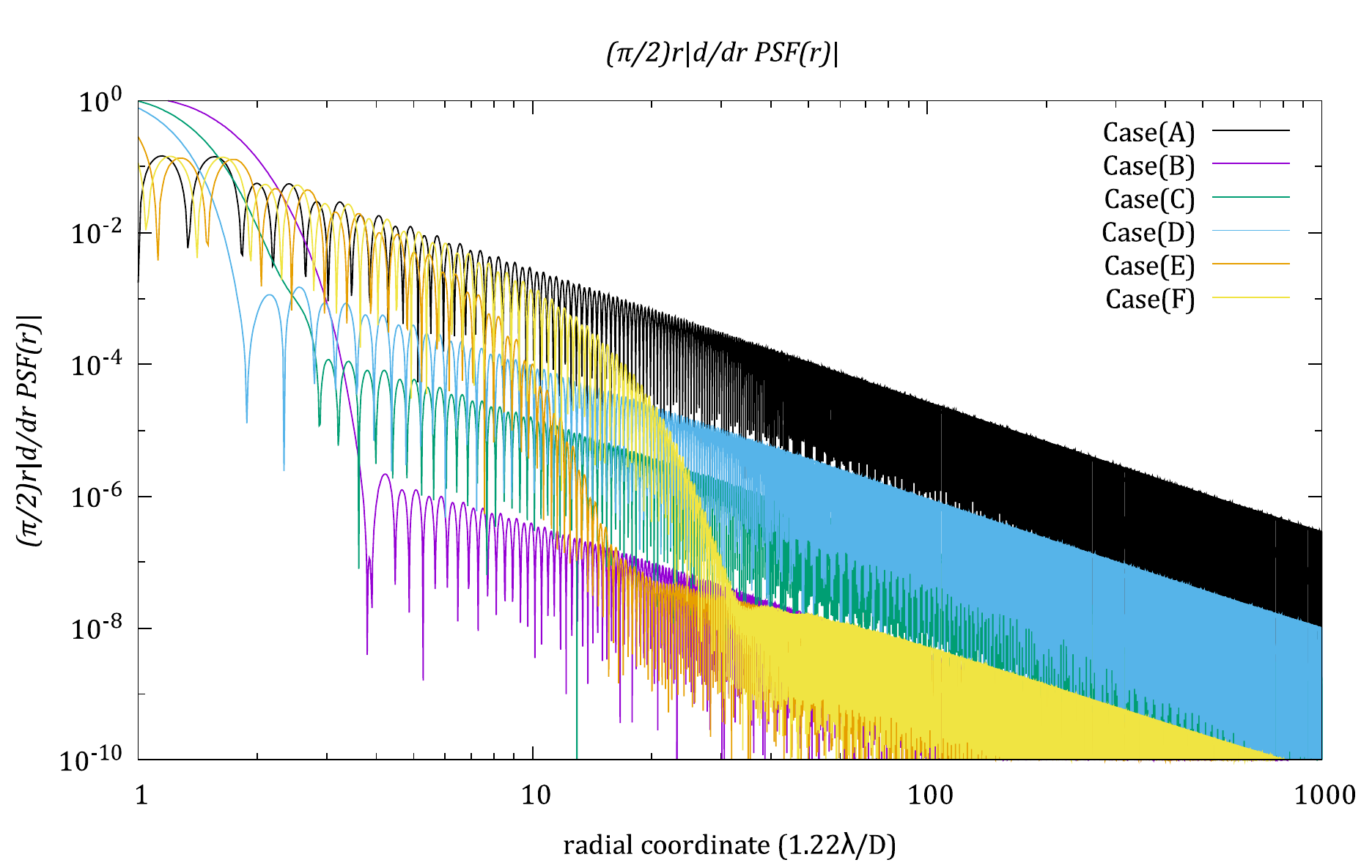}
\caption{\label{PSF_dash_CI_log} $\frac{\pi a}{2}\left|\protect\scalebox{0.7}{$\displaystyle \frac{\partial }{\partial r}PSF(r)\mid_{r=a} $}\right|$ for circular isotropic apodizing masks}
\end{center}
\end{figure}

\begin{figure}[H]
\begin{center}
\includegraphics[width=80mm]{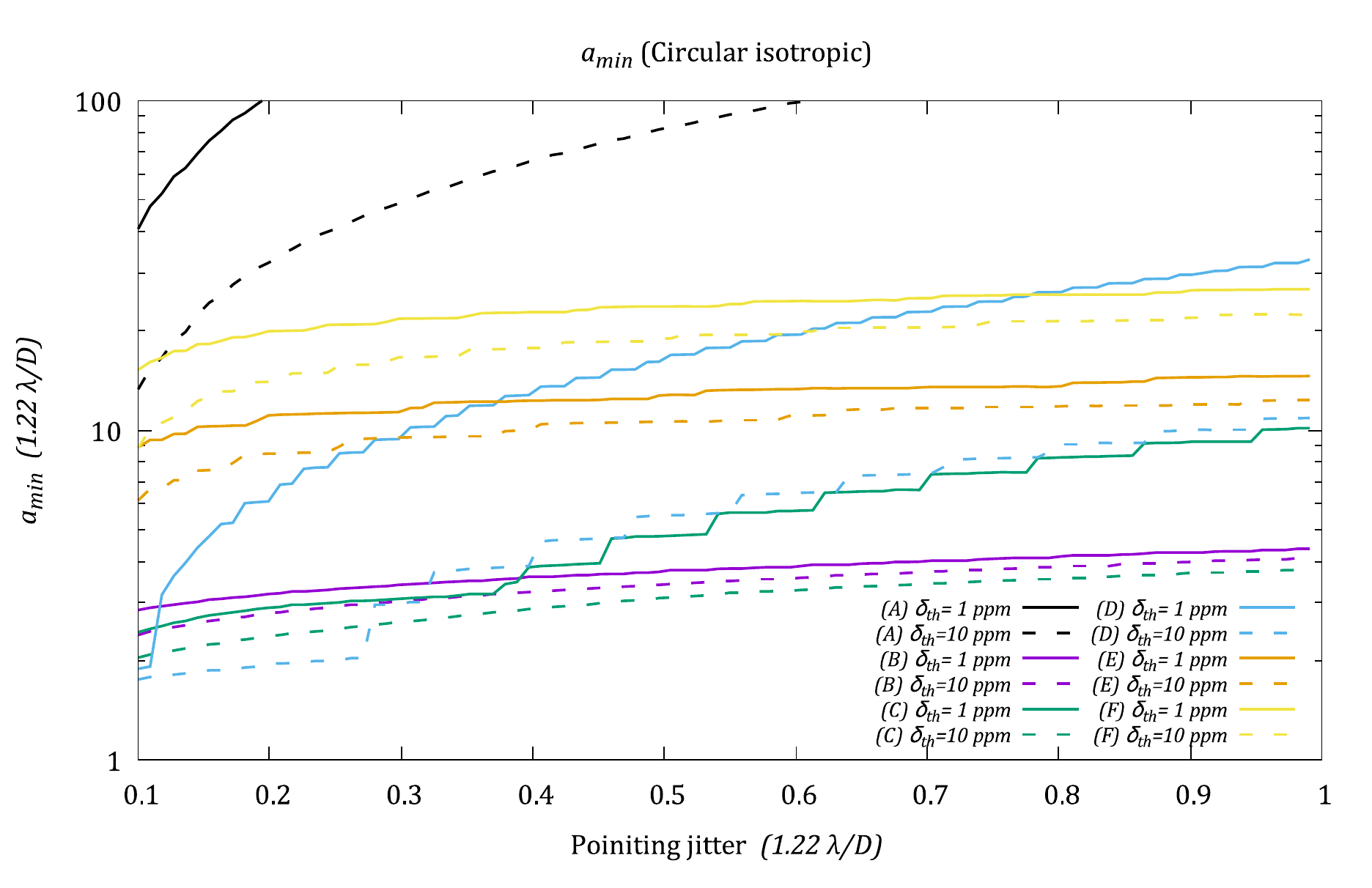}
\caption{\label{a_min_CI_log}$a_{min}$ for circular isotropic apodizing masks}
\end{center}
\end{figure}

\begin{figure}[H]
\begin{center}
\includegraphics[width=80mm]{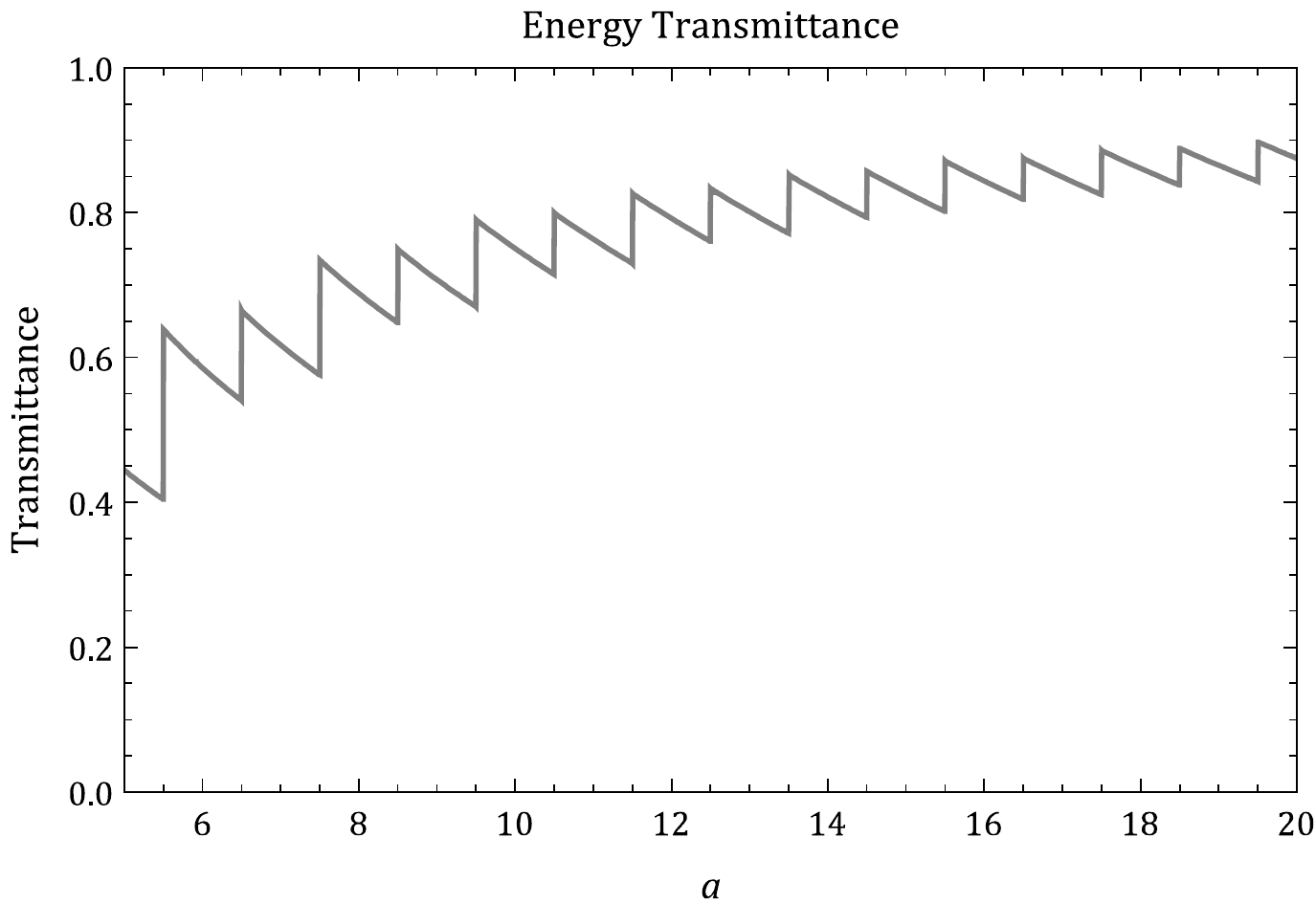}
\caption{\label{ET_1D}Energy transmittance of 1-D block-shaped mask as a function of $a$}
\end{center}
\end{figure}

\begin{figure}[H]
\begin{center}
\includegraphics[width=80mm]{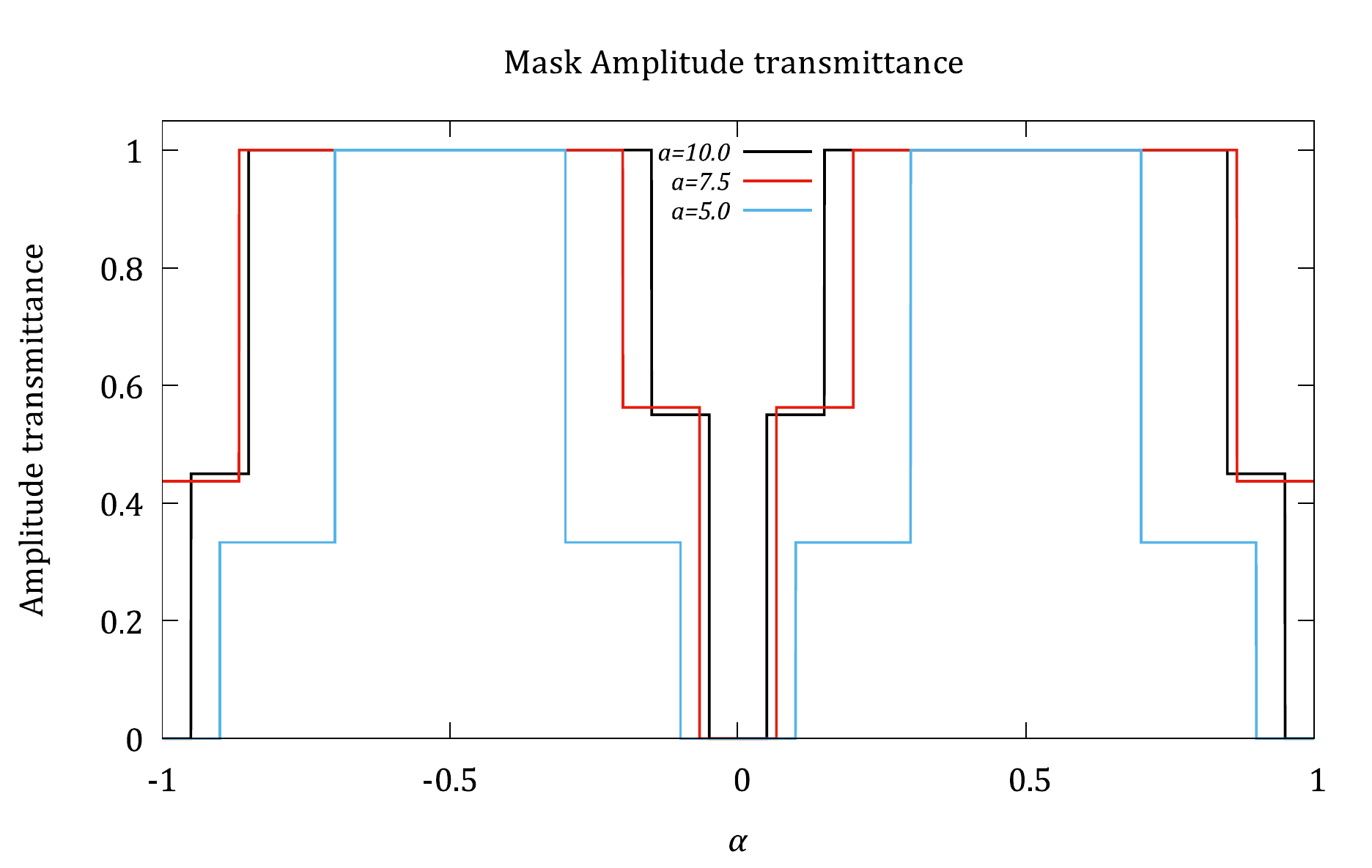}
\caption{\label{P1DN}Pupil functions of 1-D block-shaped masks}
\end{center}
\end{figure}

\begin{figure}[H]
\begin{center}
\includegraphics[width=80mm]{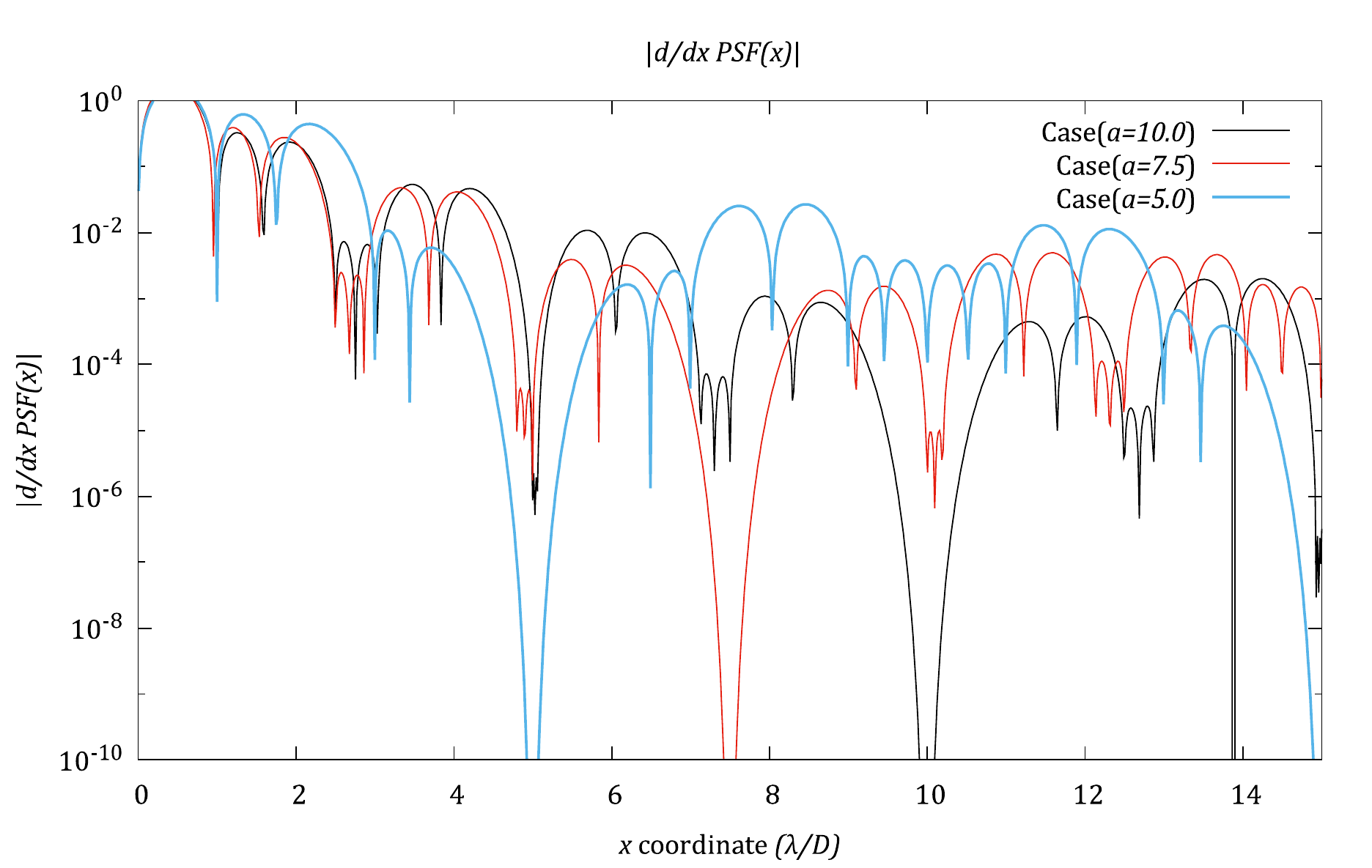}
\caption{\label{PD1DN}$\left|\frac{d}{dx}\protect\scalebox{0.7}{$\displaystyle PSF(x)\mid_{x=a}$}\right|$ for 1-D block-shaped masks}
\end{center}
\end{figure}

\begin{figure}[H]
\begin{center}
\includegraphics[width=80mm]{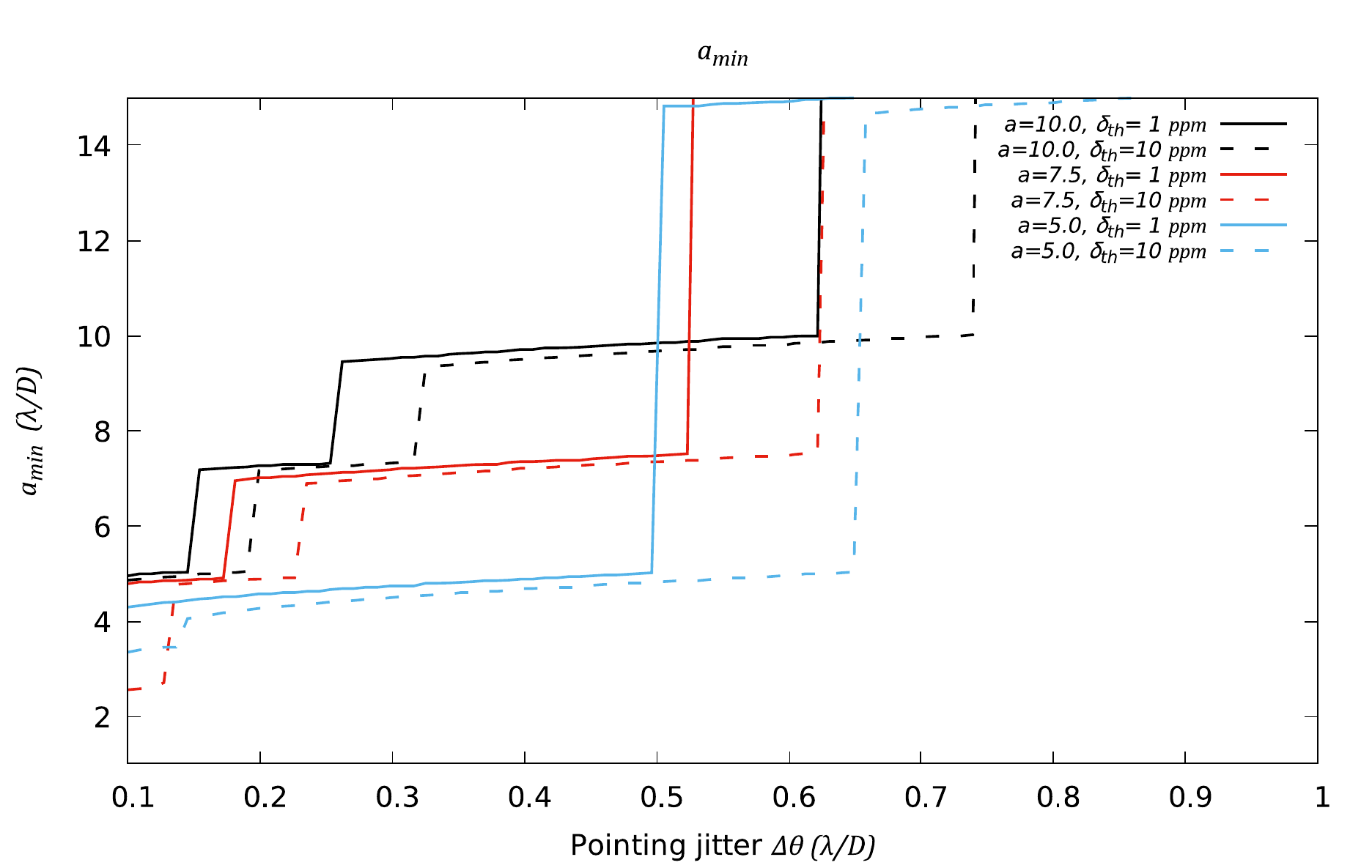}
\caption{\label{a_min_1D_N}$a_{min}$ for 1-D block-shaped masks}
\end{center}
\end{figure}

\begin{figure}[H]
\begin{center}
\includegraphics[width=80mm]{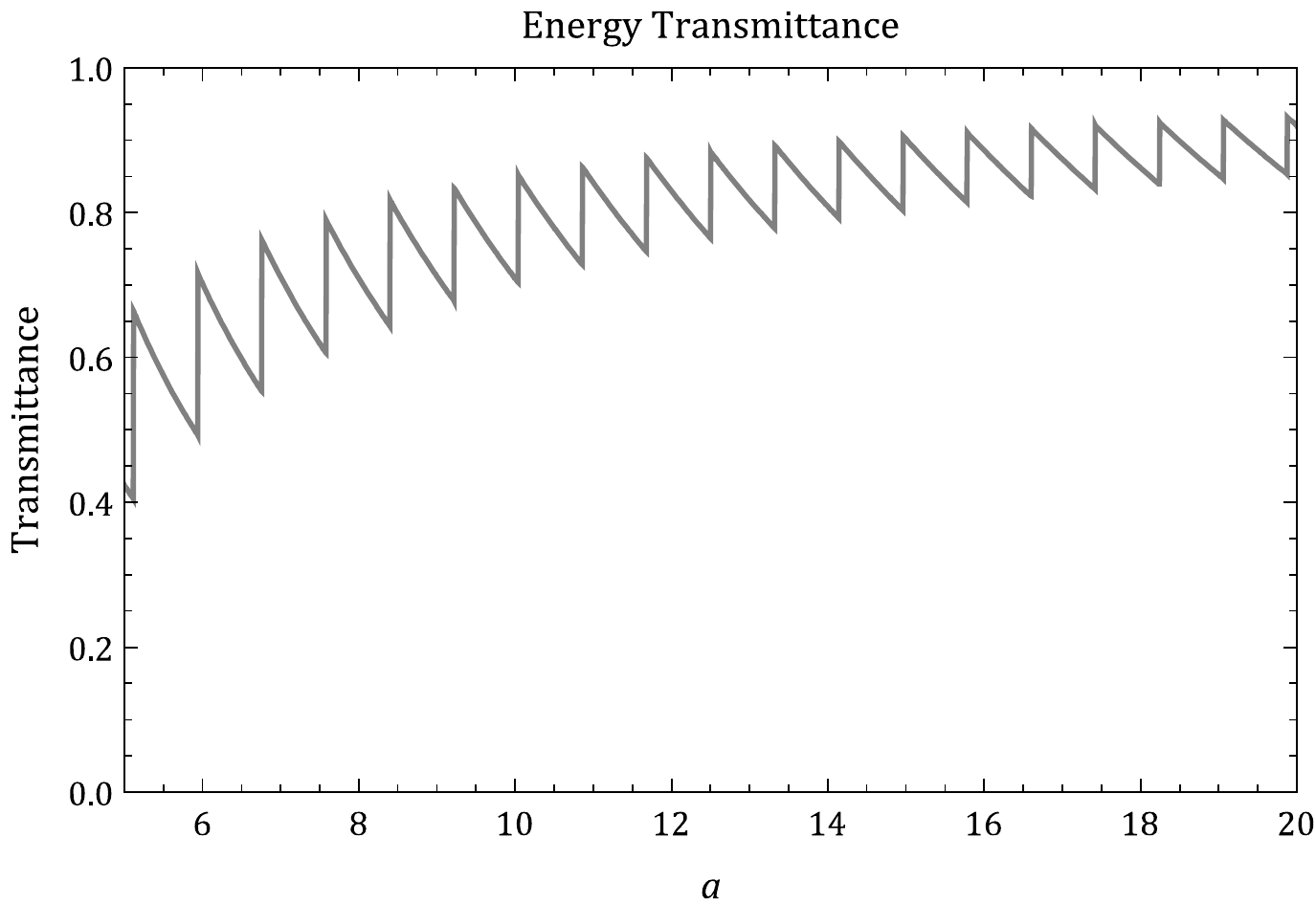}
\caption{\label{ET_CI}Energy transmittance of circular isotropic block-shaped mask as a function of $a$}
\end{center}
\end{figure}

\begin{figure}[H]
\begin{center}
\includegraphics[width=80mm]{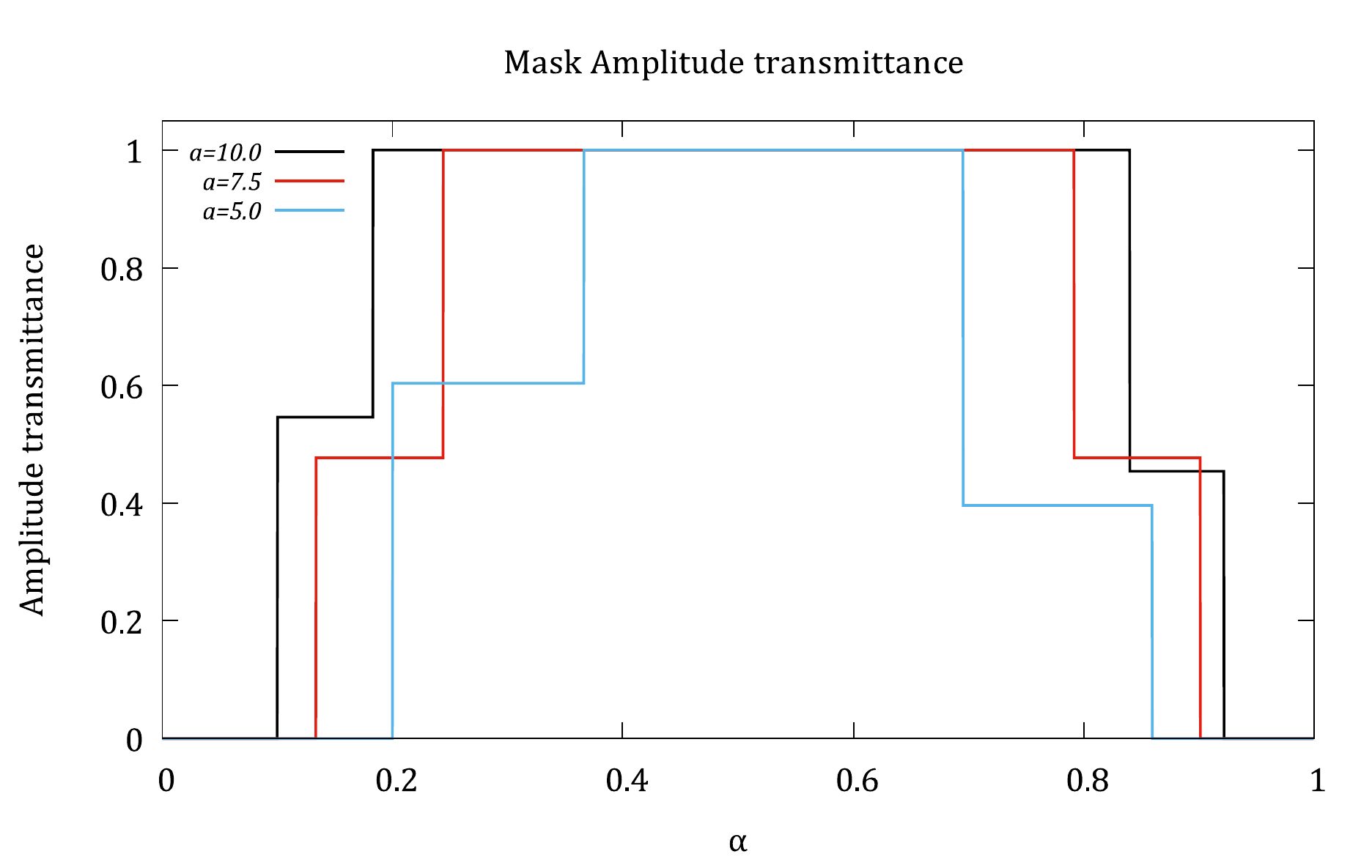}
\caption{\label{PCIN}Pupil functions of circular isotropic block-shaped masks}
\end{center}
\end{figure}

\begin{figure}[H]
\begin{center}
\includegraphics[width=80mm]{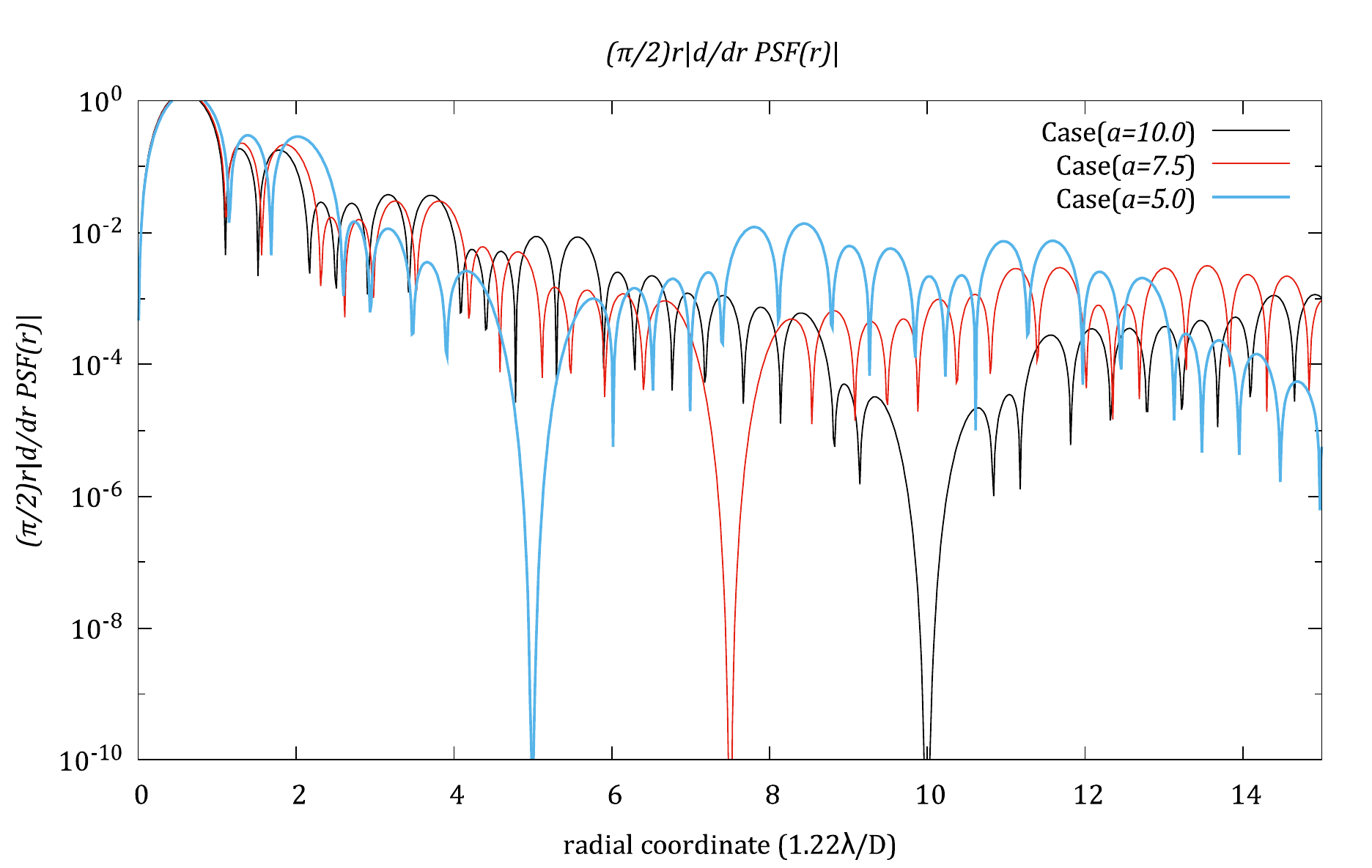}
\caption{\label{PDCIN}$\frac{\pi a}{2}\left|\protect\scalebox{0.7}{$\displaystyle \frac{\partial }{\partial r}PSF(r)\mid_{r=a} $}\right|$ for circular isotropic block-shaped masks}
\end{center}
\end{figure}

\begin{figure}[H]
\begin{center}
\includegraphics[width=80mm]{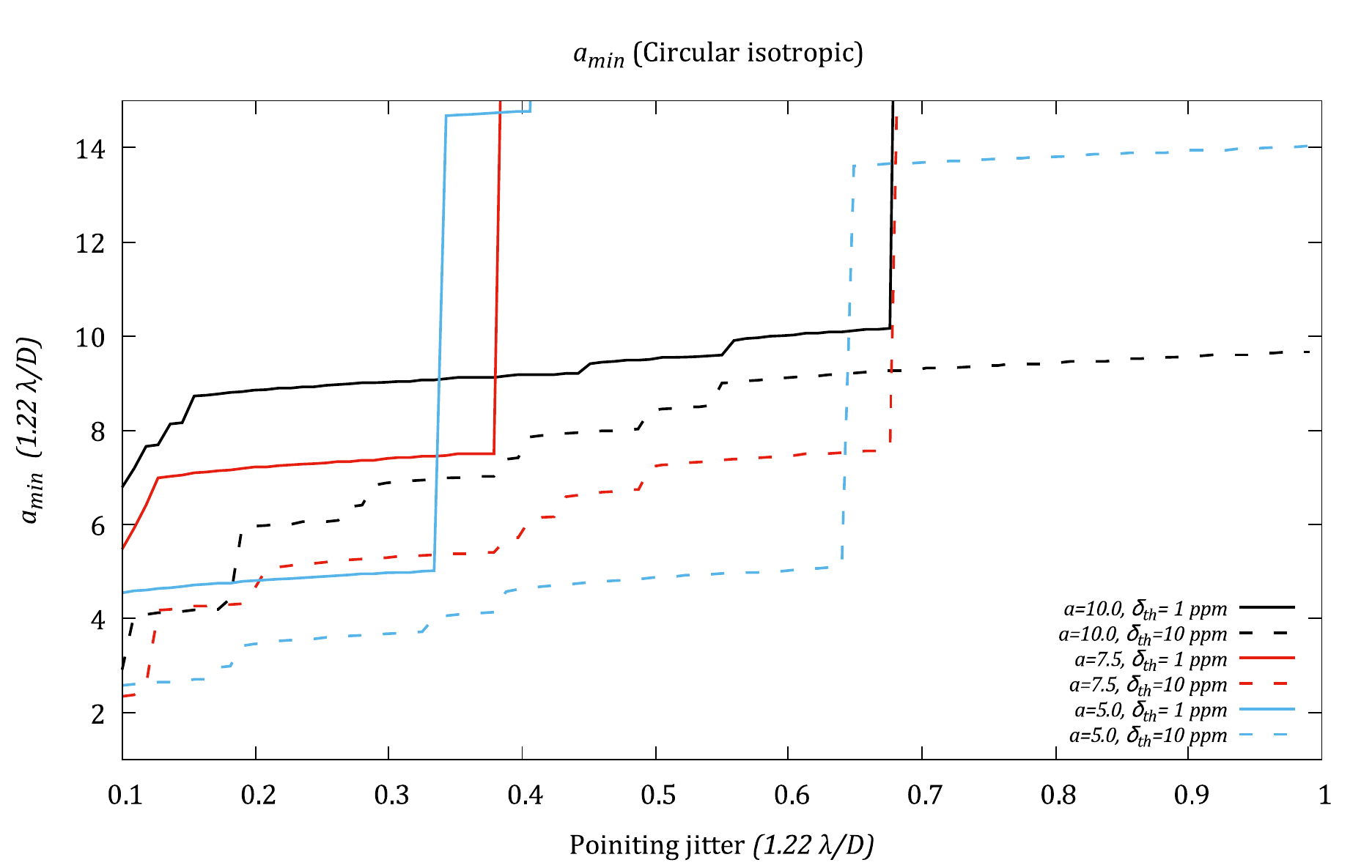}

\caption{\label{a_min_CI_N} $a_{min}$ for circular isotropic block-shaped masks}
\end{center}
\end{figure}

\begin{figure}[H]
\begin{center}
\includegraphics[width=80mm]{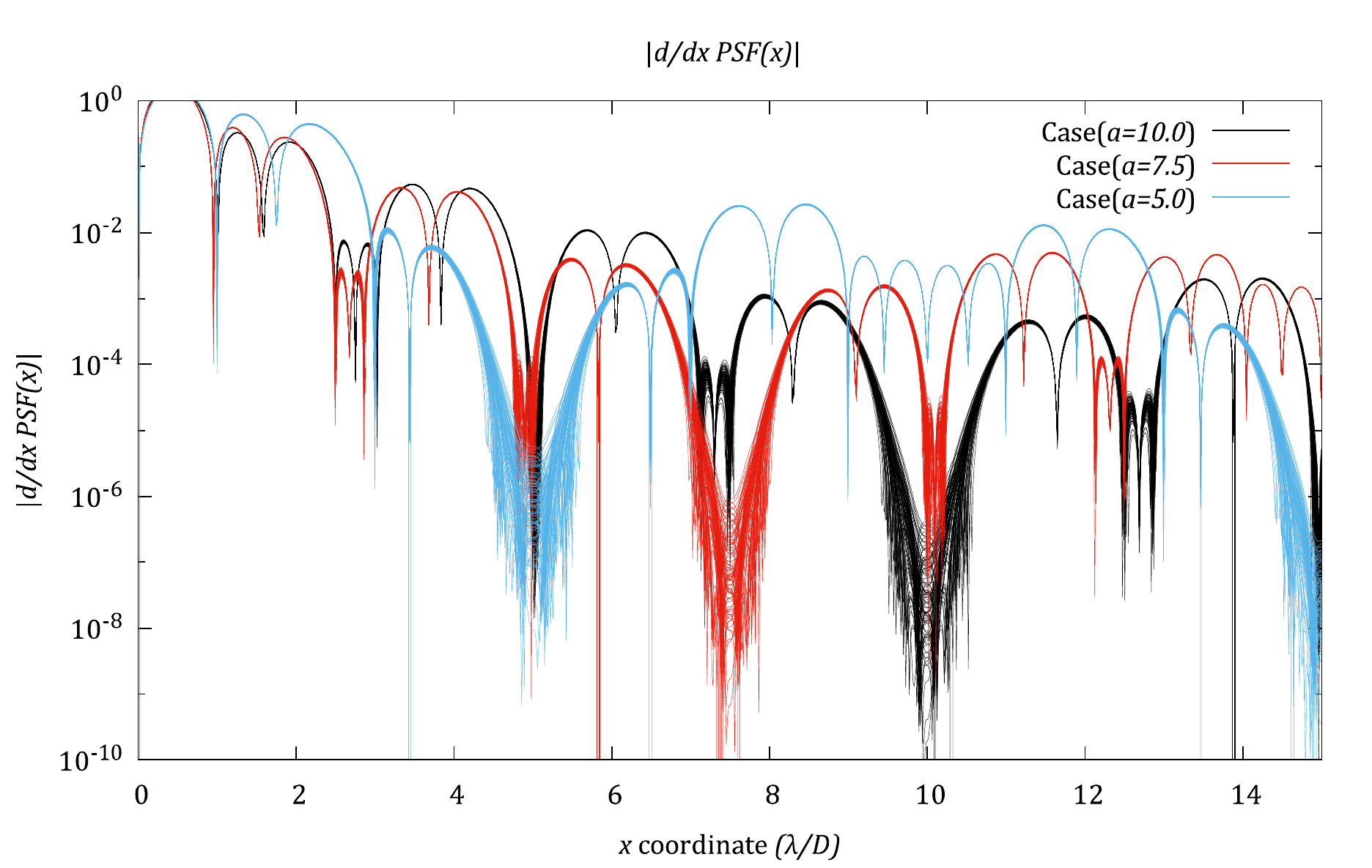}
\caption{\label{PD1D_tor}$\left|\frac{d}{dx}\protect\scalebox{0.7}{$\displaystyle PSF(x)\mid_{x=a}$}\right|$ for 1-D block-shaped masks with transmittance deviation}
\end{center}
\end{figure}

\begin{figure}[H]
\begin{center}
\includegraphics[width=80mm]{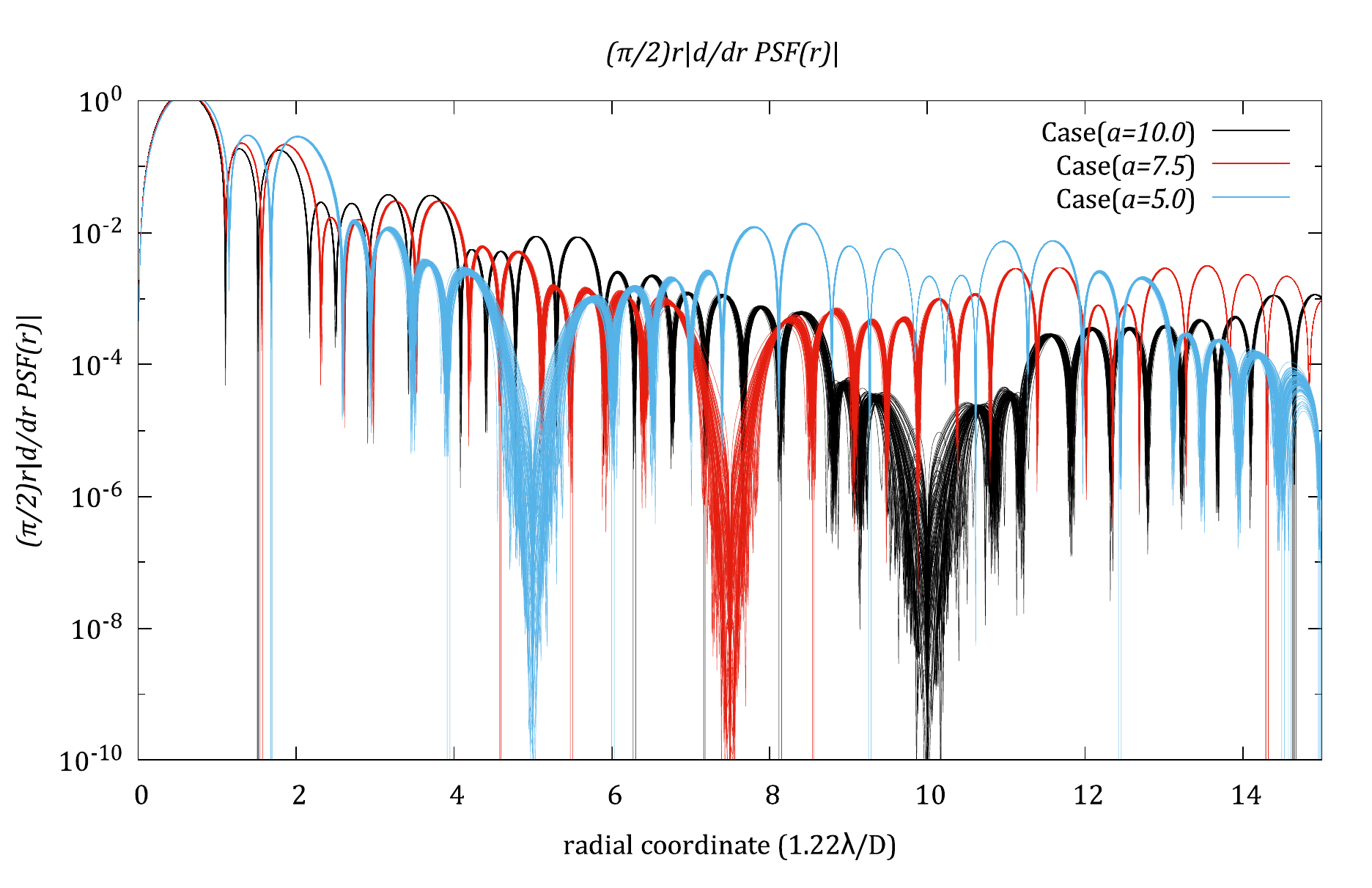}
\caption{\label{PDCI_tor}$\frac{\pi a}{2}\left|\protect\scalebox{0.7}{$\displaystyle \frac{\partial }{\partial r}PSF(r)\mid_{r=a} $}\right|$ for circular isotropic block-shaped masks with transmittance deviation}
\end{center}
\end{figure}

\begin{figure}[H]
\begin{center}
\includegraphics[width=80mm]{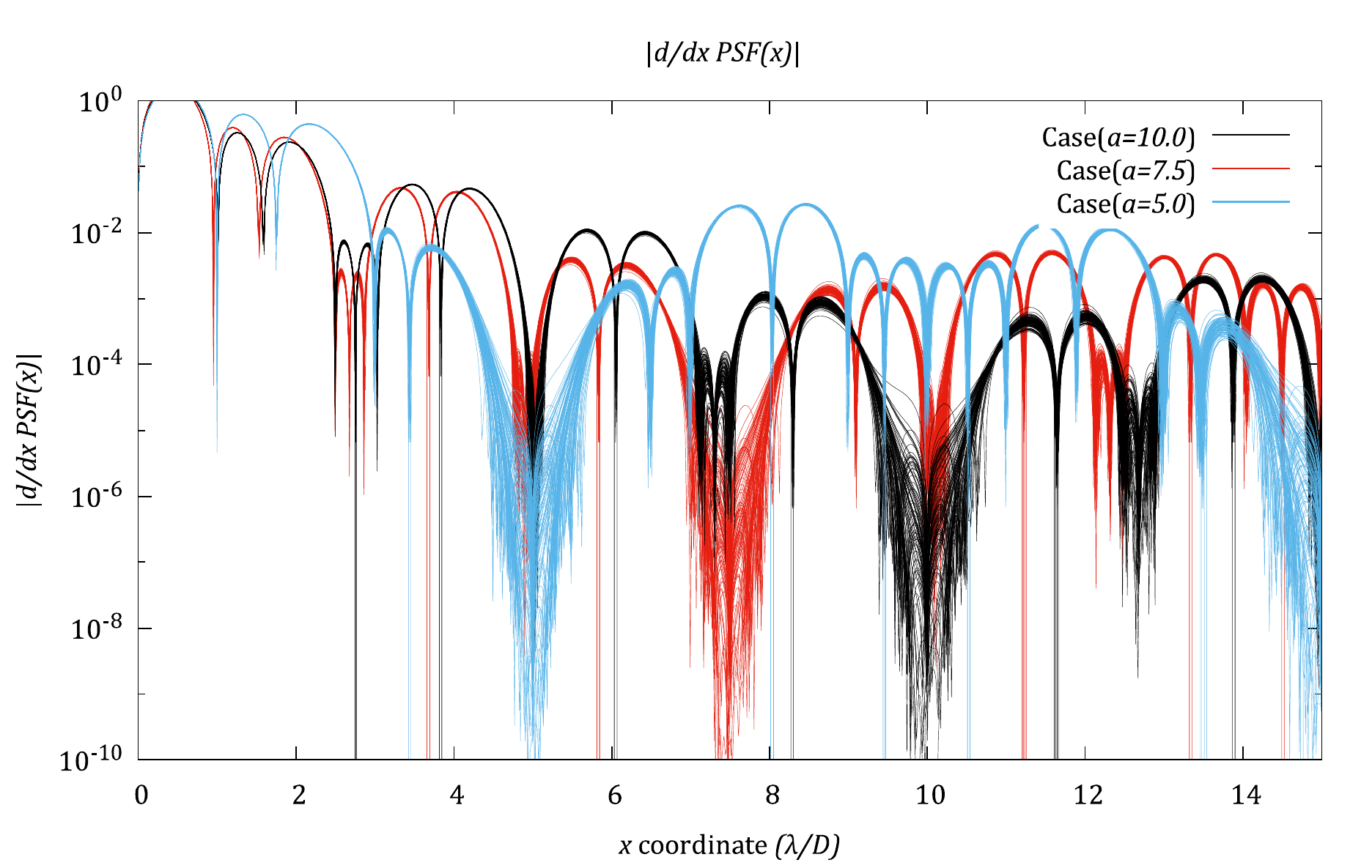}
\caption{\label{PD1D_divi}$\left|\frac{d}{dx}\protect\scalebox{0.7}{$\displaystyle PSF(x)\mid_{x=a}$}\right|$ for 1-D block-shaped masks with  positional deviation of transmittance changing point}
\end{center}
\end{figure}

\begin{figure}[H]
\begin{center}
\includegraphics[width=80mm]{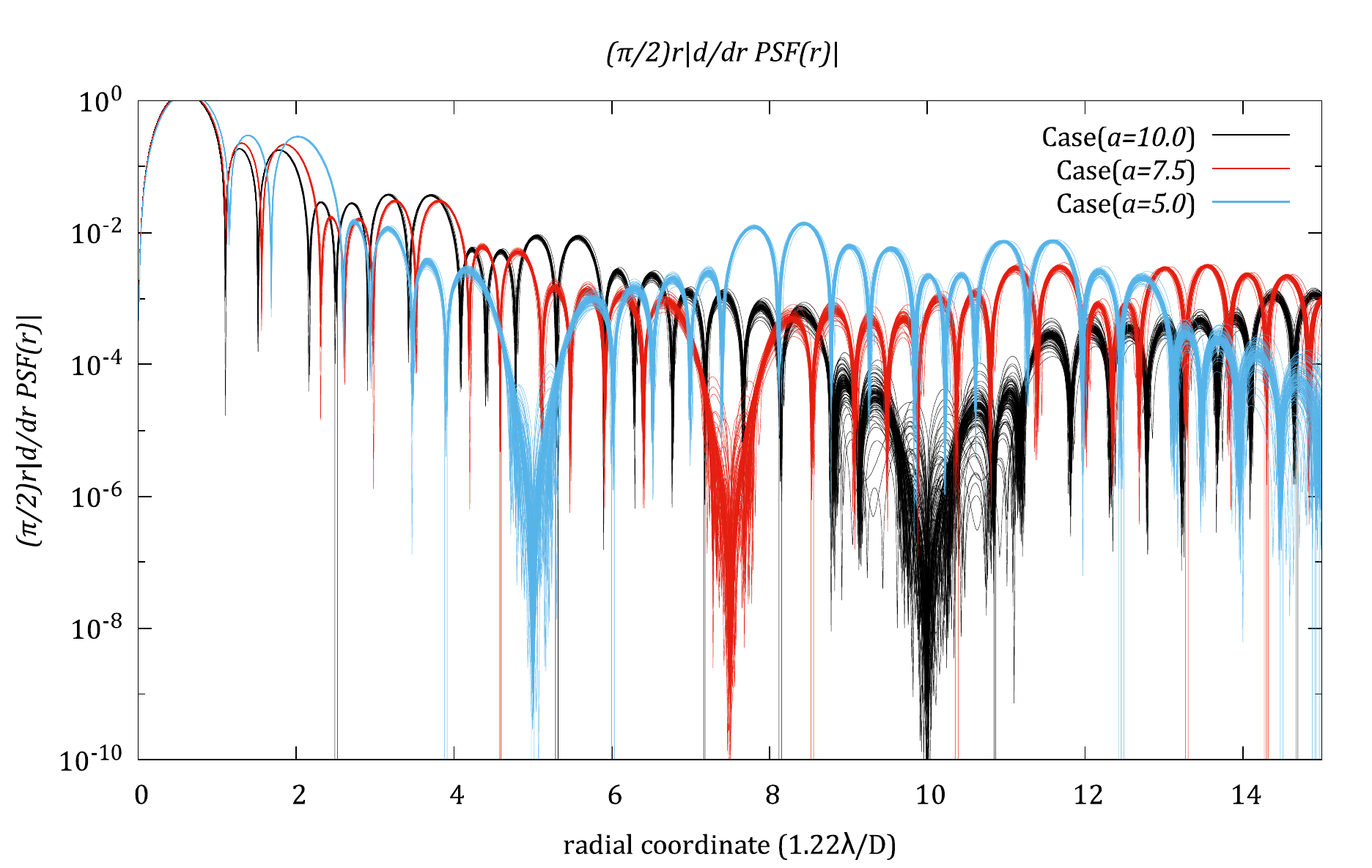}
\caption{\label{PDCI_divi}$\frac{\pi a}{2}\left|\protect\scalebox{0.7}{$\displaystyle \frac{\partial }{\partial r}PSF(r)\mid_{r=a} $}\right|$ for circular isotropic block-shaped masks with positional deviation of transmittance changing point}
\end{center}
\end{figure}

\end{document}

%% file: PMSTE_abstract.tex
Spectrophotometric stability, which is crucial in the spectral characterization of transiting exoplanets, is affected by photometric variations arising from field-stop loss in space telescopes with pointing jitter or primary mirror deformation. This paper focuses on a new method for removing slit-loss or field-stop-loss photometric variation through the use of a pupil mask. Two types of pupil function are introduced: the first uses conventional (e.g., Gaussian or hyper-Gaussian) apodizing patterns; whereas the second, which we call a block-shaped mask, employs a new type of pupil mask designed for high photometric stability. A methodology for the optimization of a pupil mask for transit observations is also developed. The block-shaped mask can achieve a photometric stability of $10^{-5}$ for a nearly arbitrary field-stop radius when the pointing jitter is smaller than approximately $0.7  \lambda/D $ and a photometric stability of $10^{-6}$ at a pointing jitter smaller than approximately $0.5 \lambda/D $.   The impact of optical aberrations and mask imperfections upon mask performance is also discussed.

%% file: PMSTE_introduction.tex
Spectral characterization of exoplanets is important in the investigation of  planetary   
chemical and physical characteristics, such as effective temperature, surface composition, and atmospheric composition (Charbonneau et al. 2008, Swain et al 2008, Belu et al. 2011, Burrows 2014a). In particular, secondary eclipsing in the mid-infrared wavelength range provides opportunities to detect emission spectra from temperate terrestrial planets in the habitable zone (Kaltenegger and Traub 2009, Kaltenegger et al. 2011). However, extremely small eclipse depth of primary transmission spectroscopy and the effect of secondary eclipsing dictate that high photometric stability is required for the spectral characterization of terrestrial planets (see Table \ref{depth}). 

 High photometric stability is fundamental in obtaining reliable transit observations. Space telescopes can be used to obtain extremely highly stable spectrophotometry results because they are not affected by intensity fluctuation arising from atmospheric turbulence and telluirc absorption. However, even from the space, photometric stability is limited not only by photon noise and the activity of host stars but also by systematic drivers within the measuring instrument (Barron et al. 2006, Seager et al. 2008, Deming et al. 2009, Beichman et al. 2014), including variation in light loss at the entrance slit of the spectrometer arising from telescope pointing jitter (Smith et al. 2007,  Crossfield et al.  2011), which is also known as  silt loss. An entrance slit is generally used in astronomical spectroscopy to limit the field of view in front of the dispersive element to avoid contamination from the spectra of background stars, galaxies, sky background, and extended light sources such as  zodiacal light. (Edwards et al. 1988, K$\ddot{\mathrm{u}}$mmel et al.  2009). The entrance slit can also be used to control  wavelength resolution. Recently, Matsuo et al. (2016) proposed a highly stable spectrophotometry method, which is immune to telescope pointing jitter, called densified pupil spectrometry. This technique employs a field stop to prevent contamination from background stars; however, it has been noted that variation of the partial occultation of point spread function (PSF) caused by the presence of the field stop limits the stability of raw observational data.

Because photometric variation arising from the slit- or field-stop-loss depends on the shape of PSF, the photometric variation can be reduced thorough the use of pupil masks, such as those used for direct imaging of extra-solar planets (Slepian 1965, Kasdin et al. 2003), to perform amplitude modulation (Figure \ref{LAYOUT}).

This paper develops a methodology for evaluating the performance of pupil masks for use in the highly stable spectrophotometry of exoplanets and uses this methodology to  propose a new type of pupil modulation.  
In Section 2, we derive an approximate formula for the simple evaluation of the photometric variation arising from the partial occultation of the PSF by the field stop, and introduce performance metrics for the pupil modulation patterns created by pupil masks. We then evaluate the performance of several conventional apodized pupil masks. In Section 3, we propose several new pupil mask patterns that depress variation of the PSF at arbitrary positions corresponding to the edge of the field stop.
 In Section 4, we discuss the viability of block-shaped masks by evaluating the impact of optical aberrations and mask imperfections upon mask performance.

%% file: PMSTE_theory_header.tex
In this section, we formulate a  general theory of performance evaluation of pupil masks for the highly stable spectrophotometry of transiting exoplanets. Following some preparatory discussion, we derive an analytical expression for the simple estimation of field-stop-loss variation arising from pointing jitter and develop some performance metrics for pupil functions used in highly stable spectrophotometry. Basing on these metrics, we evaluate the performance of several conventional pupil apodizing masks.

%% file: PMSTE_theory_preparation.tex
To begin with, we will briefly review the mechanism of Fraunhofer diffraction of one-dimensional (1-D) and circular isotropic apertures.   

We first consider the 1-D diffraction integral shown in Equation (\ref{19}), which is obtained from the separation of variables for the Fraunhofer diffraction integral of a rectangular aperture (Born and Wolf 1999): 
\begin{equation}
f(\zeta )=2\int^{d}_{0}d\xi A(\xi) \ cos\left( \frac{2\pi \zeta \xi}{\lambda f} \right)  ,
\label{19}
\end{equation}
where $\lambda$ is the wavelength, $f$ is the distance between the pupil plane and the focal plane, $d$ is the pupil radius, $\xi$ and $\zeta$ are coordinates on the
 pupil and image planes, respectively, and $A(\xi)$ is the pupil function. Setting $A(\xi)$ as uniformly unity on the pupil, using the dimensionless values $x=\zeta/\frac{\lambda f}{2 d_{max} },\ \alpha = \xi/d_{max},\ \mathrm{and}\ t=d/d_{max}$, transforms Equation (\ref{19}) to
\begin{equation}
g(x)=2d_{max}\int^{t}_{0}d\alpha \ cos\left( \pi x \alpha \right) =2d_{max}\times \frac{\sin(\pi t x)}{\pi x}.
\label{sinc}
\end{equation}

Next, the diffraction integral of a circular isotropic pupil function can be expressed by the following 0-th order Hankel transform, in which the constant factor is left out (as it is not necessary in this application):
\begin{equation}
f(\gamma )=\int^d_0 d\rho 2\pi \rho A(\rho) J_0 \left( \frac{2\pi \gamma \rho}{\lambda f} \right) ,
\label{33}
\end{equation}
where $\rho$ and $\gamma$ are the radial coordinates on the pupil and image planes, respectively. When $A(\rho)$ is uniformly unity on the pupil, the nondimensionalization of Equation (\ref{33}) by defining the dimensionless values $r=\gamma/\frac{(j_{1,1}/(2\pi))\lambda f}{ d_{max} },\  \alpha = \rho/d_{max},\mathrm{and}\ t=d/d_{max}$ produces 
 \begin{equation}
g(r)=d_{max}^2\int^t_0 d\alpha 2\pi \alpha J_0 \left( j_{1,1}r\alpha \right)=\frac{2\pi}{j_{1,1}}d_{max}^2\times \frac{t J_1(j_{1,1}tr)}{r} ,
\label{jinc}
\end{equation}
where $j_{n,m}$ is the m-th null point of the n-th Bessel function and $j_{1,1}/(2\pi)\approx 0.61$. 

The symbol notations used throughout this paper are compiled in Table \ref{notation}. 

%% file: PMSTE_theory_1-D.tex
By assuming a PSF normalized by the central intensity as a function of $x$, the photometric variation, $V$, can be expressed by
\begin{equation}
V=\frac{\left|\int^{a+\Delta \theta}_{-a+\Delta \theta}dx PSF(x)-\int^{a}_{-a}dx PSF(x)\right|}{\int^{a}_{-a}dx PSF(x)}.
\label{1}
\end{equation}
(See Table (\ref{notation}) for notation.)

Assuming that the PSF is centro-symmetric, $PSF(-x)=PSF(x)$, and the normalized pointing jitter is sufficiently less than unity, we can derive the following approximate expression for $V$ (Appendix \ref{1-D_ana}):
\begin{eqnarray}
V&=&\left|\frac{d}{dx}PSF(x)\mid _{x=a}\right|\times (\Delta \theta )^2.
\label{2}
\end{eqnarray}

Equation (\ref{2}) says that the photometric variation, $V$, is proportional to the square of the pointing jitter, and the proportionality factor can be evaluated by the absolute value of the first-order differential coefficient of the $PSF(x)$ at $x=a$.

%% file: PMSTE_theory_CI.tex
Similarly, we can derive an analytical expression for field-stop-loss variation for circular isotropic apertures.     

Writing the PSF normalized by its central intensity as a function of angular radial position on the image plane, $r=\sqrt{x^2+y^2}$, the photometric variation, $V$, can be expressed as
\begin{equation}
V=\frac{\left|\int^{a}_{-a}dy\int^{\sqrt{a^2-y^2}+\Delta \theta}_{-\sqrt{a^2-y^2}+\Delta \theta}dx PSF\left(\sqrt{x^2+y^2}\right)-\int^{a}_{0}2\pi rdr PSF(r)\right|}{\int^{a}_{0}2\pi rdr PSF(r)} .
\label{ape.8}
\end{equation}

Assuming that the normalized pointing jitter is sufficiently less than unity, we can derive the following approximate expression for $V$ (Appendix \ref{CI_ana}):
\begin{equation}
V=\frac{\pi a}{2}\left|\frac{\partial }{\partial r}PSF(r)\mid_{r=a}\right|\times (\Delta \theta )^2.
\label{2.2}
\end{equation}

This equation (\ref{2.2}) shows that the photometric variation, $V$, is proportional to the square of the pointing jitter, and the factor of proportionality is $\frac{\pi a}{2}$ times the absolute value of the first-order radial-directional-derivative coefficient of $PSF(r)$ at $x=a$.

%% file: PMSTE_theory_PerMet.tex
 In highly stable photometry, a slit or field stop is used to limit the field of view to prevent contamination from background astronomical objects such as stars or galaxies and defused light sources such as zodiacal light or sky background. Such diffused sources do not affect relative photometry except through their contributions to photon noise because the amount of  contamination from these sources does not change with telescope pointing jitter. The contamination from  galaxies is less important than that from stars. This is because the photometric error in apparent magnitude, $\Delta m$, from a partially occulted galaxy whose apparent magnitude and angular diameter are $m$ and $\theta_g$ with pointing jitter $\delta \theta $ (unnormalized) can be estimated as $\Delta m=m-2.5\delta \theta/\theta_g$. Table \ref{HUDF} indicates the relation between the V-band magnitude and the number density of galaxies observed in the Hubble ultra-deep field, and Tables \ref{M31} and Table \ref{M31over10} indicate the angular diameter and apparent magnitude of a galaxy whose size and absolute magnitude are comparable to the M31 and a galaxy 0.1 times as large as M31, respectively, for various distances. Because the apparent magnitude of an M-dwarf with an effective temperature of 3,200 K located at a distance of 10 pc is approximately 7 in N-band magnitude, 1 ppm of the flux of the star corresponds to an N-band magnitude of 22. When we assume that V-N is in the range of 1.45--6.45 (corresponding to the case in which the galactic flux per unit wavelength in the V band is $10^2$--$10^4$ times greater than that in N band.), an N-band magnitude of 22 corresponds to a V-band magnitude of 23.45--28.45. Typically, we have to consider contamination from galaxies with apparent V-band magnitudes of 25.  
Based on these considerations and the data, we conclude that the main consideration when designing the size of a field of view should be contamination from stars in the Milky Way.

Thus, a small field of view is critical for the observation of targets in dense regions of stars such as the galactic plane. In general, the observing area can be increased by reducing the radius of the field stop. However, reducing the field stop size increases occultation of the PSF side lobe of the target star, resulting in loss of light (i.e., field-stop-loss). As the field stop is located on the image plane, field-stop-loss varies with the telescope pointing jitter (Seager et al. 2008, Matsuo et al. 2016); This can lead to a photometric instability of the raw data, that is not negligible at very small eclipse depths of 1--10 ppm. 
 However, apodizing the pupil plane modifies the intensity pattern formed on the image plane from that of the original PSF (Figure \ref{LAYOUT}). Some of the resulting modified intensity patterns can lead to lower field-stop-loss variation even at small field-stop sizes, enabling highly stable photometry over small fields of view. To analyze potential apodized pupil plane intensity pattern, we introduce the following series of performance metrics for the pupil function. To obtain a certain degree of photometric  stability $\delta_{loss}$, at a given amount of pointing jitter, $\Delta \theta$, the field aperture radius must be sufficiently large. Therefore, we adopt a 
minimum working field stop angular radius, $a_{min}$, as a performance metric. The definition of $a_{min}$ is given by
\begin{eqnarray}
a_{min}(\delta_{loss}, \Delta \theta )=\min \left\{a\mid V(a+\delta a)\leq \delta_{loss}, \forall \delta a \in [-\Delta \theta ,\  \Delta \theta ] \right\} .
\end{eqnarray}
It must be noted that the normalization factor of $a_{min}$ depends on $\lambda$ and $D$.
In Figure \ref{SAvsFA}, the ratio of the observable solid angle without background star contamination to $4\pi$ is plotted as a function of a radius of field stop using the galactic stellar density model following Konishi et al. (2014) and Binney and Tremaine (1987). It is seen that, when the angular radius of the field stop is 10 arcsec, 50 \% of the overall sky can be observed without contamination. When the angular radius of the field stop exceeds 24 arcsec, no observation can be conducted without contamination.

Another performance metric is the energy transmittance of the pupil mask, $\tau$.
 
In highly stable photometry, photon noise tends to be dominant.
 Since we require many photons to enhance the signal-to-photon-noise ratio in highly stable photometry, the standard deviation of the photon-number fluctuation tends to be great. That must be greater than the reciprocal of the eclipse depth that is required to be detected with 1$\sigma$ reliability. For example, detection of a 1-ppm eclipse with 1$\sigma$ reliability requires $10^{12}$ photons; thus,  photon-number fluctuation is greater than $10^{6}$, whereas exemplary values of readout noise and dark current are $10e^{-}\ \mathrm{read}^{-1}$ and $1e^{-}\ \mathrm{pix}^{-1}s^{-1}$ (the data from the Near-Earth Object Wide-Field Infrared Survey Explorer Reactivation Mission (NEOWISE) (Cutri et al. 2015)) .
 Thus, the main design consideration in the energy transmittance of the pupil mask is how many photons are needed.
Increasing photon noise per photon number as a result of degradation of the energy transmittance limits observable region along the line of sight. 
Note that the maximum distance along the line of sight under a given set of conditions is proportional to $\sqrt{\tau}$; correspondingly, we adopt both $\tau$ and $a_{min}(\delta_{loss}, \Delta \theta)$ as performance of the pupil function.

In addition, we must note that the pupil masks slightly change the spectral resolution, $R=\lambda/\Delta\lambda$, by modifying the pupil image and the PSF of the target star. 
Since pupil masks tend to depress the light on the outer region of the pupil, the spectral resolution tends to be less in conventional-spectroscopy cases and grater in densified-pupil-spectroscopy case.

%% file: PMSTE_theory_ApoMas.tex
We first consider a Gaussian function mask as a basic apodizing pattern (Slepian 1965, Weisstein 2015):
\begin{equation}
P(x)=\exp \left( - \frac{x^2}{\sigma^2} \right) ,
\end{equation}
where $x$ is the position in the pupil plane normalized by the pupil radius and $\sigma$ is a parameter. A hyper-Gaussian function mask with higher energy transmittance than a  conventional Gaussian function mask is given by 
\begin{eqnarray}
P(x)= \left\{
\begin{array}{l}
1 \ \ \ \ \ (|x|\leq a') \\
\exp \left( -\left( \frac{|x-a'|}{b'}\right) ^n\right) \ \ \ \ \ \left( a' < |x| \leq 1 \right) 
\end{array}
\right. ,
\end{eqnarray}
where $a'$, $b'$, and $n$ are parameters of the hyper-Gaussian function.  In Figure \ref{PUPIL_MASK}, we show the amplitude transmittances for apodized pupil masks with the parameters compiled in Table \ref{mask}. We evaluated the performance of the 1-D rectangle and that of the circular isotropic pupil functions with the pupil functions transformed through the Fraunhofer diffraction integral into the PSF on the image plane (Li and  Wolf 1984). The energy transmittance, $\tau$, for each case is shown in Table \ref{mask}.

\subsubsection{1-D apertures}
We now examine the performance metrics in terms of the 1-D pupil function. Figure \ref{PSF_dash_1D_log} shows\ $\left|\frac{d}{dx}\scalebox{0.7}{$\displaystyle PSF(x)\mid_{x=a}$}\right|$\  for cases (A)-(F) from Table \ref{mask}, whereas Figure \ref{a_min_1D_log} shows \  $a_{min}$\ for\ $\delta_{loss}=10^{-5} \ \mathrm{and}\  10^{-6}$.
For apodizing masks (Cases (B)-(F)), \ $\left|\frac{d}{dx}\scalebox{0.7}{$\displaystyle PSF(x)\mid_{x=a}$}\right|$\ is obviously smaller than in the case of no mask (Case (A)).
The Gaussian masks, with smaller $\sigma$ (Case (B)), depress \ $\left|\frac{d}{dx}\scalebox{0.7}{$\displaystyle PSF(x)\mid_{x=a}$}\right|$\ significantly at smaller $x$, whereas a hyper-Gaussian mask, with larger energy transmittance (Case (F)), has larger \ $\left|\frac{d}{dx}\scalebox{0.7}{$\displaystyle PSF(x)\mid_{x=a}$}\right|$\ at smaller $x$.
Therefore, $a_{min}$ for Gaussian mask is much smaller than in the case of no mask.
For example, when $\Delta \theta= 0.5$ and $\delta_{loss}= 10^{-6}$, $a_{min}$ in case (B) is around 4 while that in case (A) is approximately 90.
 
\subsubsection{Circular isotropic apertures}
We next consider the performance in terms of the circular isotropic pupil functions. Figure \ref{PSF_dash_CI_log} shows\  $\frac{\pi a}{2}\left|\scalebox{0.7}{$\displaystyle \frac{\partial }{\partial r}PSF(r)\mid_{r=a} $}\right|$\  for cases (A)-(F), whereas Figure \ref{a_min_CI_log} shows\  $a_{min}$\ for\ $\delta_{loss}= 10^{-5} \ \mathrm{and}\  10^{-6}$.
For each mask, the trends seen in $V=\frac{\pi a}{2}\left|\scalebox{0.7}{$\displaystyle \frac{\partial }{\partial r}PSF(r)\mid_{r=a} $}\right|$ and $a_{min}$ are roughly similar to those for 1-D apertures; however, although the performances of the hyper-Gaussian mask is nearly identical in both cases, the performances of Gaussian masks with $\sigma= 0.5\ \mathrm{and}\ 0.6$ (Case (C) and (D), respectively) are not as good in the circular isotropic case.

%% file: PMSTE_new_header.tex
In this section, we evaluate masks with PSF intensity variation of approximately zero only in the vicinity of the edge of the field stop (over a certain width range) and show that these are close to ideal for use in highly stable spectrophotometry. To do this, we successfully construct pupil functions with PSF intensity variation in the vicinities of arbitrary positions that are close to zero mathematically and derive these functions in analytical form. In the following, we will describe the  mathematical expression (the derivation for which is compiled in Appendix \ref{both_MF}) and evaluate the performance of these new pupil mask designs in transit spectrophotometric observation.

%% file: PMSTE_new_ME_1-D.tex
We now show the mathematical expression for the new type of masks with PSF-intensity variation of approximately zero only at $x=a$ in the 1-D case.

As building blocks, we present the following pupil functions, $P(\delta ,a;\alpha )$, (Appendix \ref{1-D_MF}.):
\begin{equation}
P(\delta ,a;\alpha )= \mathrm{Rect}\biggl[\frac{\alpha}{B(\delta ,a)}\biggr] - \mathrm{Rect}\biggl[ \frac{\alpha}{L(\delta ,a)} \biggr] ,
\end{equation}
where 
 \begin{eqnarray}
\mathrm{Rect}[\alpha ] = \left\{
\begin{array}{l}
1 \ \ \ \ \ \left(|\alpha |\leq 1 \right) \\
0 \ \ \ \ \ \left( 1 < |\alpha | \right) 
\end{array}
\right.
\end{eqnarray}
and
\begin{eqnarray}
L(\delta,a )&=&\frac{\delta}{2a} \ \ \ \ (\delta=1,3 )\nonumber \\
B(\delta,a )&=&L+2/a\times \left\lfloor \frac{a}{2}-\frac{\delta}{4} \right\rfloor .
\end{eqnarray}
By superposing the pupil functions for $\delta=1$ and $\delta=3$, we obtain that of our novel block-shaped mask, $Q(a;\alpha )$, as (Appendix \ref{1-D_MF}.):
\begin{equation}
Q(a;\alpha )=W_1(a)P(1,a;\alpha )+W_3(a)P(3,a;\alpha ) ,
\end{equation}
where the weighting factors are determined by 
\begin{equation}
W_{\delta}(a)=\frac{(B(\delta ,a)^2 - L(\delta ,a)^2)}{(B(1,a)^2 - L(1,a)^2)+(B(3,a)^2 - L(3,a)^2)} \ \ \ \ (\delta=1,3 ).
\end{equation}

This function, $Q(a;\alpha )$, is designed such that the zeroth-, first-, second- and third-order differential coefficients for the PSF at $x=a$ are zero.

%% file: PMSTE_new_ME_CI.tex
We can also derive masks with a PSF-intensity variation of approximately zero only at $r=a$ in the circular isotropic case.

We use the following pupil functions, $P(\delta ,a;\alpha )$ (Appendix \ref{CI_MF}.):
\begin{equation}
P(\delta ,a;\alpha )= \mathrm{Circ}\biggl[\frac{\alpha}{B(\delta ,a)}\biggr] - \mathrm{Circ}\biggl[ \frac{\alpha}{L(\delta ,a)} \biggr] ,
\label{CITOP}
\end{equation}
where 
\begin{eqnarray}
\mathrm{Circ}[\alpha ]= \left\{
\begin{array}{l}
1 \ \ \ \ \ \left(0 \leq \alpha \leq 1 \right) \\
0 \ \ \ \ \ \left( 1 <\alpha  \right) 
\end{array}
\right .
\end{eqnarray}
and
\begin{eqnarray}
L(\delta ,a)&=&\frac{j_{1,1+\delta}}{j_{1,1}a} \ \ \ \ (\delta = 0,1) \nonumber \\
B(\delta ,a)&=&\max_{n \in N} \left\{\frac{j_{1,2n+\delta}}{j_{1,1}a} | \frac{j_{1,2n+\delta}}{j_{1,1}a}\leq 1 \right\} \ \ \ \ (\delta = 0,1)
\end{eqnarray}
By superposing these pupil functions, a block-shaped mask for the circular isotropic case, $Q(a;\alpha )$, is expressed as follows (Appendix \ref{CI_MF}.):
\begin{equation}
Q(a;\alpha )=W_0(a)P(0,a;\alpha ) + W_1(a)P(1,a;\alpha )
\end{equation}
where the weighting factors are determined by 
\begin{equation}
W_{\delta}(a )= \frac{|J_2(j_{1,1}a B(\delta,a )) B(\delta,a )^2-J_2(j_{1,1}a L(\delta ,a ))L(\delta ,a )^2| }{\sum_{\delta ' =0}^1|J_2(j_{1,1}a B(\delta ' ,a )) B(\delta ' ,a )^2-J_2(j_{1,1}a L(\delta ' ,a ))L(\delta ' ,a )^2|} .
\label{CIEND}
\end{equation}

Here, $Q(a;\alpha )$ is designed such that zeroth-, first-, and second-order differential coefficients for the PSF at $r=a$ are zero.

%% file: PMSTE_new_perf_1-D.tex
The $Q(a;\alpha)$ described above is a set of functions with a parameter $a$. Figure \ref{ET_1D} indicates the energy transmittance of the masks $\tau(a)=\int^1_0 Q(a;\alpha)^2d\alpha$ as a function of $a$. 
The energy transmittance roughly increases with $a$, and the function has sharp local maxima when $a$ is half-integer.
Figure \ref{P1DN} indicates the pupil functions at $a=5.0,\ 7.5,\  \mathrm{and}\  10$. 
We call this new type of mask block-shaped mask because of the appearance of the resulting  pupil functions.
Figure \ref{PD1DN} and \ref{a_min_1D_N} shows $\left|\frac{d}{dx}\scalebox{0.7}{$\displaystyle PSF(x)$}\right|$ and $a_{min}$, respectively,  under the same conditions.
$\left|\frac{d}{dx}\scalebox{0.7}{$\displaystyle PSF(x)$}\right|$ is depressed at $x=a$ and $a_{min}$ is equal to $a$ for values of $\Delta \theta$ smaller than around 0.5 ($\delta_{loss} = 10^{-6} $) and 0.7 ($\delta_{loss} = 10^{-5} $), although the performance is limited at larger $\Delta \theta$ because the corresponding width of the depression around $x=a$ is not very  wide.   

%% file: PMSTE_new_perf_CI.tex
Figure \ref{ET_CI} shows the energy transmittance of masks $\tau(a)=\frac{1}{\pi}\int^1_0 Q(a;\alpha)^2 2\pi \alpha d\alpha$ as a function of $a$.
As in the 1-D case, the energy transmittance increases with $a$, while the function has sharp local maxima at $a$ is in $\left\{\frac{j_{1,n}}{j_{1,1}} \mid n \in N  \right\}$.
Figure \ref{PCIN} shows the pupil functions at  $a=5.0, 7.5,\  \mathrm{and}\ 10$. 
These pupil functions shapes differ only slightly from those in the 1-D case.
Figure \ref{PDCIN} and \ref{a_min_CI_N} show $\frac{\pi a}{2}\left|\scalebox{0.7}{$\displaystyle \frac{\partial }{\partial r}PSF(r) $}\right|$ and $a_{min}$, respectively, under the same conditions. 
The width of depression in $V$ is narrower than in the 1-D case because the third-order  differential coefficients are not zero, unlike the 1-D case. However, the circular isotropic aperture performance is about as good as in the 1-D case because the baseline of $V$ is depressed more than it is in the 1-D case. 
These results indicate that the block-shaped mask can enable us highly stable spectrophotometry in a space telescope with a high degree of pointing jitter.

%% file: PMSTE_discussion_header.tex
The above discussion has assumed ideal optics. Now we discuss the impact of optical aberrations and mask imperfections.

%% file: PMSTE_discussion_IOA_header.tex
The pupil masks based on diffraction phenomena described above are only effective when the PSF width is almost diffraction limited. Hence our working hypothesis requires almost diffraction-limited PSF, but, simultaneously, it must be noted that other diffuser-like pupil mask technologies (e.g., diffuser-like pupil mask in Wide-field Infrared Camera (WIRC) at the 200-inch Hale telescope at Palomar Observatory) can make non-diffraction-limited PSF spread out over many pixels and stable. 
When we conduct observing using  ground-based telescopes without adaptive optics (AO) with pupil masks based on diffraction phenomena, the pupil must be smaller enough than Fried's coherence length (Fried. 1965) to allow seeing.  Hence, we assume ground-based observation using AO or pupils smaller than Fried's coherence length, or space-based observation with a balloon, airplane, or artificial satellite, where either little wavefront distortion and low-order optical aberration exist or else they are well compensated.

%% file: PMSTE_discussion_IOA_HOA.tex
To test the impact of high order wavefront error, we use the following model as the power-spectral density (PSD) of  high-order phase aberration following Traub and Oppenheimer, 2010:
\begin{equation}
PSD(k)=\frac{3\sqrt{3}}{4\pi ^2 k_0^2}\frac{1-S}{1+\left( \frac{k}{k_0} \right) ^3},
\end{equation}
where $S$ is the Strehl ratio calculated from the root mean square (RMS) of the wavefront error,  $k$ is the absolute value of the wave number of the wavefront fluctuation, and $k_0=10\ \mathrm{cycle}/\mathrm{pupil} $ is a model parameter. The unit of $k$ is $\mathrm{cycle}/\mathrm{pupil}$, and the unit of $PSD(k)$ is $\mathrm{radian}^2/(\mathrm{cycle}/\mathrm{pupil})^2$.
Now let us assume wavefront error is sufficiently smaller than 1 radian. Defining $PSF_0(r)$ and $PSF_{\Psi}(r)$ as the PSF with and without the wavefront aberration, respectively, leads to
\begin{equation}
PSF_{\Psi}(r)=PSF_0(r)+PSF_0(r)\ast PSD(r),
\end{equation}
where $\ast$ is the 2-D - convolution operator with respect to $x$ and $y$, which satisfies  $r=\sqrt{x^2+y^2}$. Due to convolution theorem, defining $\mathcal{F}$ as the 2-D - Fourier-transform operator with respect to $x$ and $y$ then leads to
\begin{equation}
PSF_{\Psi}(r)=PSF_0(r)+\mathcal{F}^{-1}[\mathcal{F}[PSF_0(r)]\times \mathcal{F}[ PSD(r)]].
\label{abe}
\end{equation}
$\mathcal{F}[PSF_0(r)](k)$ in Equation (\ref{abe}) is the auto correlation function (ACF) of the pupil function without aberration (i.e., the mask-amplitude transmittance). Because $\mathcal{F}[ PSD(r)](k)$ has relatively large values only in the region where $k$ is small compared to the k-direction width of the ACF of the pupil function without aberration, the approximation $\mathcal{F}[PSF_0(r)](k)\approx \mathcal{F}[PSF_0(r)](0)=\pi \tau$ is valid in Equation (\ref{abe}). Thus, the equation simply becomes
\begin{equation}
PSF_{\Psi}(r)\approx PSF_0(r)+ \pi \tau PSD(r).
\end{equation}
This representation depends on mask type only by $\tau$. In other words, the impact of high-order aberration is almost unchanged by mask type. Hence, $\frac{V}{\Delta\theta^2}$ is limited by
\begin{equation}
\frac{\pi a}{2}\left | \frac{\partial}{\partial r}PSF_{\Psi}(r)\mid _{r=a}\right | \leq  \frac{\pi a}{2}\left | \frac{\partial}{\partial r}PSF_{0}(r)\mid _{r=a}\right | +\frac{\tau \pi ^2 a}{2}\left | \frac{\partial}{\partial r}PSD(r)\mid _{r=a}\right |.
\end{equation}
The wavefront-aberration-contribution term is
\begin{equation}
\frac{\tau \pi ^2 a}{2}\left | \frac{\partial}{\partial r}PSD(r)\mid _{r=a}\right |=\frac{9\sqrt{3}\tau a}{8000}\frac{1-S}{\left( 1+\left( \frac{a}{10}\right)^3 \right) ^2},
\end{equation}
and, for example, the values of  $V$ of the block-shaped mask at $a=5.0,\ 7.5,\ 10$ are less than $3\times 10^{-3}(1-S)\Delta\theta^2,\ 6\times 10^{-3}(1-S)\Delta\theta^2,\ 4\times 10^{-3}(1-S)\Delta\theta^2$, respectively. When we assume that $S=90\%$,\ $\Delta \theta = 0.1$, the values of  $V$ become less than $3\ \mathrm{ppm},\ 6\ \mathrm{ppm},$ and $4\ \mathrm{ppm}$, respectively.

%% file: PMSTE_discussion_IMI_header.tex
Block-shaped  masks are expected to be manufactured in a similar way to the well-known neutral density (ND) Filter. In particular, it is advisable to base the design on reflective ND filters to achieve constancy of transmittance over a wide wavelength range. Reflective ND filters are made of a transparent medium coated in a thin metallic layer to reduce transmittance. Both glass substrates coated with chrome for visible light and ZnSe substrates coated with nickel for mid-infrared light (2-16 micron) are commercially available. Additionally, ND filters whose transmittance change with the position on them like block-shaped mask also has been commercially available. Thus, these technologies can be directly applied to manufacturing block-shaped masks. Testing the transmittance distribution will be done using a laser beam whose shape has been well investigated by a knife-edge test.
Keeping the above discussion in mind, here we evaluate the impact of the transmittance deviation and the positional deviation of the transmittance-changing point upon the performance of the block-shaped mask. 

%% file: PMSTE_discussion_IMI_TD.tex
The region where the energy transmittance should be 0 \% of the  block-shaped mask is made easily using a medium that is perfectly opaque under absorption or reflection. We can also  redefine the transmittance of the mask substrate to 100 \%. Thus, we only have to consider the transmittance deviation on the intermediate-transmittance region, that is, the region on witch  the energy transmittance is neither 0 \% nor 100 \%. 
Hence, in Figures \ref{PD1D_tor} and \ref{PDCI_tor}, we show the resulting 100 perturbed samples of $\frac{V}{\Delta\theta^2}$ as thin lines assuming that the amplitude transmittances on the intermediate-transmittance regions (4 intervals for the 1-D case and 2 rings for the circular isotropic case) independently deviate from the designed value following the normal distribution with $1\ \mathrm{standard\ deviasion} =1\%$.
In the 1-D case, $V\leq 10\Delta\theta^2\ \mathrm{ppm}$ at $a=5.0$, and $V\leq 1\Delta\theta^2\ \mathrm{ppm}$ at $a=10$. 
In the circular isotropic case, though the depression width is less than in the 1-D case, $V\leq 10\Delta\theta^2\ \mathrm{ppm}$  at  $a=5.0,\ 7.5,\ \mathrm{and}\ \ 10$. 

%% file: PMSTE_discussion_IMI_PD.tex
In the derivation of the block-shaped mask (Appendix \ref{both_MF}), we tuned the positions where mask transmittance change discontinuously such that $\frac{V}{\Delta\theta^2}$ would have a lower value in the vicinity of the desired position. Hence, mask performance seems to be relatively sensitive to the positions at which the mask transmittance changes discontinuously. To estimate this impact, in Figures \ref{PD1D_divi} and \ref{PDCI_divi}, we showed the resulting 100 perturbed samples of $\frac{V}{\Delta\theta^2}$ as thin lines, assuming that the distance from the center of the pupil to the boundaries at which the mask transmittance changes discontinuously (4 points in 1-D case and 4 circles in circular isotropic case) independently deviate from the designed value following a normal distribution with $1\ \mathrm{standard\ deviasion} =0.1\%$ of pupil radius.
In the 1-D case, $V\leq 10\mathrm{-}100\times \Delta\theta^2\ \mathrm{ppm}$ at $a=5.0,\ 7.5,\ \mathrm{and}\ \ 10$.
In the circular isotropic case, $V\leq 100\Delta\theta^2\ \mathrm{ppm}$ at $a=5.0,\ 7.5,\ \mathrm{and}\ \ 10$.

%% file: PMSTE_discussion_OU.tex
Block-shaped masks may be useful for standard photometry and polarimetry. This is because the smaller the aperture for aperture photometry is, the less the photometry result is affected by contamination from foreground or background sources.  In the same manner, such masks is useful to reduce contamination in pupil photometry with smaller field stops. Additionally, it could be useful for coronagraphic applications such as direct imaging of exoplanets provided that the separation of the parent star and the target has been well investigated. 

%% file: PMSTE_conclusion.tex
This paper presented a method for removing photometric instability arising from slit- or field-stop-loss coupled to telescope pointing jitter. This research was undertaken with the goal of attaining highly stable spectrophotometry of transiting exoplanets. As the spectrum depths of primary and secondary eclipses for terrestrial planets around late-type stars are very small (corresponding to 2-10 ppm), it is fundamental in undertaking reliable transit observations to ensure the high photometric stability of the raw data.

 We first derived a simple analytical expression for the photometric variation of slit- and field-stop-loss caused by pointing jitter. Basing on this, we developed a methodology for the optimization of pupil masks used for spectrophotometric transit observations in terms of two performance metrics: the mask energy transmittance, $\tau$, and the minimum field-stop radius available for a given photometric stability and pointing jitter, $a_{min}$. These two quantities are useful because they are respectively associated with the observable distance along the line of sight and the observable sky region under given viewing circumstances. By using this methodology, we evaluated the performances of Gaussian and hyper-Gaussian masks.

We also developed a new type of pupil function suitable for highly stable spectrophotometry, which we call the block-shaped mask. Our design guideline for the block-shaped mask was to construct a PSF with an intensity variation in the vicinity of the edge of the field stop as close to zero as possible. For circular isotropic apertures, we constructed the pupil functions, $Q(a;\alpha )$, expressed by Equations (\ref{CITOP})-(\ref{CIEND}).

$Q(a;\alpha )$ is designed so that zeroth-, first-, and second-order differential coefficients for the PSF at $R=a$, as represented by the Fraunhofer diffraction integral of $Q(a;\alpha )$ are zero and the energy transmittance is as large as possible.
This mask can achieve a photometric stability of $10^{-5}$ at a  field-stop radius $a$ for a  pointing jitter roughly smaller than $0.854\lambda/D$ (i.e., the product of $0.7\ \mathrm{and\ the\ Fraunhofer\ diffraction\ limit}\ 1.22\lambda/D $ and a photometric stability of $10^{-6}$ at a field-stop radius  $a$ for a pointing jitter smaller than roughly $0.61\lambda/D $(or the product of $0.5$ and the corresponding Fraunhofer limit).

We  also discussed the impact of optical aberration and mask imperfection upon mask performance. The impact of optical aberration only weakly depends upon mask type. The typical value of  $V$ is a several hundred times $\Delta\theta^2(1-\mathrm{(Strehl\ ratio)})$ ppm. The mask performance is sensitive to the positional deviation of the transmittance-changing point rather than the transmittance deviation. When the positional deviation of this point is less than 0.1 \% of the pupil radius and all other factors are ideal, we can guarantee that $V \leq 100\Delta\theta^2\ \mathrm{ppm}$.

%% file: PMSTE_acknowledgement.tex
We are sincerely grateful to Dr. Tomoyasu Yamamuro for discussion on feasibility of application of these apodized pupil masks to densified pupil spectrometer. We also appreciate Dr. Mihoko Konishi for kind instruction on estimation of background stars based on the galactic star model.

%% file: PMSTE_appendix_DAEFV_1D.tex
By assuming a PSF normalized by the central intensity as a function of $x$, the photometric variation, $V$, can be expressed as
\begin{equation}
V=\frac{\left|\int^{a+\Delta \theta}_{-a+\Delta \theta}dx PSF(x)-\int^{a}_{-a}dx PSF(x)\right|}{\int^{a}_{-a}dx PSF(x)}.
\label{ape.1}
\end{equation}
Because the denominator in Equation (\ref{ape.1}) is nearly unity when $a$ is sufficiently large, Equation (\ref{ape.1}) can be approximated as 
\begin{equation}
V=\left|\int^{a+\Delta \theta}_{-a+\Delta \theta}dx PSF(x)-\int^{a}_{-a}dx PSF(x)\right|.
\end{equation}
Defining $G(x)$ as a primitive function of $PSF(x)$ then leads to
\begin{equation}
\int^{a+\Delta \theta}_{-a+\Delta \theta}dx PSF(x)=G(a+\Delta \theta )-G(-a+\Delta \theta ).
\label{ape.3}
\end{equation}
Taylor expanding the right hand side around $a\  \mathrm{and}\  -a$ in Equation(\ref{ape.3}) produces
 \begin{equation}
\int^{a}_{-a}dx PSF(x)+\sum^{\infty}_{n=1} \frac{\left( PSF^{(n-1)}(a)-PSF^{(n-1)}(-a)\right)}{n!}(\Delta \theta )^n.
\label{ape.4}
\end{equation}
Because the assumption of centro-symmetry leads to $PSF^{(n)}(x)=(-1)^{n}PSF^{(n)}(-x)$, the terms corresponding to $(\Delta \theta)^{n}$ with odd $n$ vanish and the summation in Equation (\ref{ape.4}) becomes
 \begin{eqnarray}
\sum^{\infty}_{n=1} \frac{\left( PSF^{(n-1)}(a)-PSF^{(n-1)}(-a)\right)}{n!}(\Delta \theta )^n \nonumber \\ =2 \sum^{\infty}_{k=1} \frac{PSF^{(2k-1)}(a)}{(2k)!}(\Delta \theta )^{2k}.
\end{eqnarray}
Ignoring the terms corresponding to $(\Delta \theta)^{n}$ ($n=2k$) equal to or larger than $n=4$ for approximation, the photometric variation $V$ can be approximated as
\begin{eqnarray}
V&=&\left|\int^{a+\Delta \theta}_{-a+\Delta \theta}dx PSF(x)-\int^{a}_{-a}dx PSF(x)\right|\nonumber \\ &=&\left|\frac{d}{dx}PSF(x)\mid _{x=a}\right|\times (\Delta \theta )^2.
\end{eqnarray}

%% file: PMSTE_appendix_DAEFV_CI.tex
By further developing the 1-D calculation above, we can derive an analytical expression for field-stop-loss variation of circular isotropic apertures.     

Writing the PSF normalized by its central intensity as a function of angular radial position on the image plane, $r=\sqrt{x^2+y^2}$, the photometric variation $V$ can be expressed as
\begin{equation}
V=\frac{\left|\int^{a}_{-a}dy\int^{\sqrt{a^2-y^2}+\Delta \theta}_{-\sqrt{a^2-y^2}+\Delta \theta}dx PSF\left(\sqrt{x^2+y^2}\right)-\int^{a}_{0}2\pi rdr PSF(r)\right|}{\int^{a}_{0}2\pi rdr PSF(r)} .
\label{ape.8}
\end{equation}
As the denominator in Equation(\ref{ape.8}) does not differ significantly from unity if $a$ is sufficiently large, we can approximate $V$ as 
\begin{equation}
V=\left|\int^{a}_{-a}dy\int^{\sqrt{a^2-y^2}+\Delta \theta}_{-\sqrt{a^2-y^2}+\Delta \theta}dx PSF\left(\sqrt{x^2+y^2}\right)-\int^{a}_{0}2\pi rdr PSF(r)\right|.
\label{ape.13}
\end{equation}
Defining $G_y(x)$ as a primitive function of $PSF\left(\sqrt{x^2+y^2}\right)$ with respect to $x$ then leads to
 \begin{equation}
\int^{\sqrt{a^2-y^2}+\Delta \theta}_{-\sqrt{a^2-y^2}+\Delta \theta}dx PSF\left(\sqrt{x^2+y^2}\right)=G_y(\sqrt{a^2-y^2}+\Delta \theta )-G_y(-\sqrt{a^2-y^2}+\Delta \theta ).
\label{ape.14}
\end{equation}
Taylor expanding the right-hand side of Equation (\ref{ape.14}) around $\sqrt{a^2-y^2},-\sqrt{a^2-y^2}$ produces
\begin{equation}
\int^{\sqrt{a^2-y^2}}_{-\sqrt{a^2-y^2}}dx PSF\left(\sqrt{x^2+y^2}\right)+\sum^{\infty}_{n=1} \frac{\left( \frac{\partial ^{n-1}}{\partial x^{n-1}}PSF(r)\mid_{x=\sqrt{a^2-y^2}}-\frac{\partial ^{n-1}}{\partial x^{n-1}}PSF(r)\mid_{x=-\sqrt{a^2-y^2}}\right)}{n!}(\Delta \theta )^n.
\label{ape.15}
\end{equation}
The assumption of centro-symmetry causes the terms corresponding to $(\Delta \theta)^{n}$ ($n=\mathrm{odd}$) to vanish. By ignoring of the terms corresponding to  $(\Delta \theta)^{n}$ larger than n=4, Equation (\ref{ape.15}) can be approximated as follows:
\begin{equation}
\int^{\sqrt{a^2-y^2}}_{-\sqrt{a^2-y^2}}dx PSF\left(\sqrt{x^2+y^2}\right)+ \frac{\left(\frac{\partial }{\partial x}PSF(r)\mid_{x=\sqrt{a^2-y^2}}-\frac{\partial }{\partial x}PSF(r)\mid_{x=-\sqrt{a^2-y^2}}\right)}{2}(\Delta \theta )^2.
\end{equation}
Integration with respect to $\int^{a}_{-a}dy$ produces
\begin{equation}
\int^{a}_{0}2\pi rdr PSF(r)+  \frac{\pi a}{2}\frac{\partial }{\partial r}PSF(r)\mid_{r=a}\times (\Delta \theta )^2.
\end{equation}
Therefore, the variation in the photometry, $V$, can be approximated as
\begin{equation}
V=\frac{\pi a}{2}\left|\frac{\partial }{\partial r}PSF(r)\mid_{r=a}\right|\times (\Delta \theta )^2.
\end{equation}

%% file: PMSTE_appendix_MFBSM_1D.tex
As a starting point, we consider the following pupil functions, and solve the parametric conditions $B\  (B \leq 1)$ and $L\  (0 \leq L < B)$ such that the differential coefficients of zeroth and first orders of intensity patterns formed by these pupil masks are zero: 
\begin{equation}
P(\alpha )= \mathrm{Rect}\biggl[\frac{\alpha}{B}\biggr] - \mathrm{Rect}\biggl[ \frac{\alpha}{L} \biggr] ,
\end{equation}
where 
 \begin{eqnarray}
\mathrm{Rect}[\alpha ] = \left\{
\begin{array}{l}
1 \ \ \ \ \ \left(|\alpha |\leq 1 \right) \\
0 \ \ \ \ \ \left( 1 < |\alpha | \right) .
\end{array}
\right.
\end{eqnarray}
Using Equation (\ref{sinc}), the diffraction-image-amplitude distributions, or amplitude apread functions (ASFs), $g(x)$, are expressed as
\begin{equation}
g(x)=2d_{max}\times \frac{\sin(\pi B x)-\sin(\pi L x)}{\pi x} ,
\end{equation}
and $\frac{d}{dx} g(x)$ becomes
\begin{equation}
\frac{d}{dx} g(x)=2d_{max}\left(\frac{\pi B cos(\pi B x) - \pi L cos(\pi L x) }{\pi x} - \frac{\sin(\pi B x)-\sin(\pi L x)}{\pi x^2} \right) .
\end{equation}
The condition by which $g(x)$ and $\frac{d}{dx} g(x)$ must be zero at $x=a$ limits the parameters $B\  (B \leq 1)$ and $L\  (0 \leq L < B)$ as follows:
\begin{eqnarray}
L&=&\frac{\delta}{2a}+2/a\times k \ \ \ \ (k=0,1,2,...\ \ ,\delta=1,3 )\nonumber \\
B&=&L+2/a\times l \ \ \ \ (l=1,2,...\ \ ,k+l \leq \left\lfloor \frac{a}{2}-\frac{\delta}{4} \right\rfloor)\nonumber \\
& & \ \ \ \ \ \ \  \mbox{($\left\lfloor ...\right\rfloor$ means floor function.)}
\end{eqnarray}
The maximum $B$ and minimum $L$ are applied because doing so results in the largest energy transmittance. 
Using only two quantities, $\delta\  \mathrm{and}\  a$,  $B(\delta,a)\  \mathrm{and}\  L(\delta,a)$ are written as
\begin{eqnarray}
L(\delta,a )&=&\frac{\delta}{2a} \ \ \ \ (\delta=1,3 )\nonumber \\
B(\delta,a )&=&L+2/a\times \left\lfloor \frac{a}{2}-\frac{\delta}{4} \right\rfloor ,
\end{eqnarray} 
where the functions $P(\alpha )\  \mathrm{and} \  g(x)$ corresponding to these values of $B$ and $L$ are donated by $P(\delta,a;\alpha )\ \mathrm{and}\  g(\delta,a;x)$, respectively.

Using mathematical induction, the differential coefficients of $N$-th order of sinc function can be derived as follows:
\begin{equation}
\frac{d^N}{dx^N}\left( \frac{sin(\omega x)}{x}\right) = \sum^N_{K=0}\frac{N!}{(N-K)!} \frac{\omega ^{N-K}  \sin (\omega x+(N+K)\pi /2)}{x^{k+1}}
\end{equation}
Therefore, the $N$-th order differential coefficient of $g(\delta,a; x)$ at $x=a$ is expressed by
\begin{align}
\frac{d^N}{dx^N} g(\delta,a; x)|_{x=a}&= \nonumber \\ 
& \left\{
\begin{array}{l}
\frac{2d_{max}}{\pi}\sum^{N/2}_{L=0}  \frac{ (-1)^{\frac{N+\delta +2L-1}{2}} N!((\pi B(\delta ,a) )^{N-2L}-(\pi L(\delta ,a) )^{N-2L}  )}{(N-2L)! a^{2L+1}} \ \ \ \ \ (\mbox{N is even}) \\
\frac{2d_{max}}{\pi}\sum^{(N-1)/2}_{L=0}  \frac{(-1)^{\frac{N+\delta +2L}{2}} N!((\pi B(\delta ,a) )^{N-(2L+1)}-(\pi L(\delta ,a) )^{N-(2L+1)}  )}{(N-2L-1)! a^{2L+2}} \ \ \ \ \ (\mbox{N is odd}) 
\end{array}
\right. &
\end{align}
This shows that the new ASF is a superposition of $g(1,a;x)$ and $g(3,a;x)$ with  weights of $W_1(a)=\frac{(B(3,a)^2 - L(3,a)^2)}{(B(1,a)^2 - L(1,a)^2)+(B(3,a)^2 - L(3,a)^2)}$ and $W_3(a)=\frac{(B(1,a)^2 - L(1,a)^2)}{(B(1,a)^2 - L(1,a)^2)+(B(3,a)^2 - L(3,a)^2)}$, respectively:
\begin{equation}
h(a;x)=W_1(a)g(1,a;x)+W_3(a)g(3,a;x).
\end{equation}
This function's first-, second-, and third-order  differential coefficients become zero. This indicates a PSF intensity variation that is close to zero only in the vicinity of edge of the field stop. Thus, the required pupil function is derived as follows:
\begin{equation}
Q(a;\alpha )=W_1(a)P(1,a;\alpha )+W_3(a)P(3,a;\alpha ) .
\end{equation}

%% file: PMSTE_appendix_MFBSM_CI.tex
Following the previous 1-D calculation, we next consider pupil functions for circular isotropic cases:
\begin{equation}
P(\alpha )= \mathrm{Circ}\biggl[\frac{\alpha}{B}\biggr] - \mathrm{Circ}\biggl[\frac{\alpha}{L}\biggr] \ \ \  (B \leq 1 , \  0 \leq L < B) ,
\end{equation}
where
\begin{eqnarray}
\mathrm{Circ}[\alpha ]= \left\{
\begin{array}{l}
1 \ \ \ \ \ \left(0 \leq \alpha \leq 1 \right) \\
0 \ \ \ \ \ \left( 1 <\alpha  \right) 
\end{array}
\right . .
\end{eqnarray}
We limit $B$ and  $L$ as follows:
\begin{eqnarray}
L&=&\frac{j_{1,1+\delta}}{j_{1,1}a} \ \ \ \ (\delta = 0,1) \nonumber \\
B&=&\max_{n \in N} \left\{\frac{j_{1,2n+\delta}}{j_{1,1}a} | \frac{j_{1,2n+\delta}}{j_{1,1}a}\leq 1 \right\} \ \ \ \ (\delta = 0,1).
\end{eqnarray}
Based on Equation (\ref{jinc}), $g(\delta,a; r)$, corresponding to the Fraunhofer-diffraction integral of $P(\delta,a; \alpha )$, can be formulated as
\begin{equation}
g(\delta,a;r )= \frac{2\pi}{j_{1,1}}d_{max}^2\frac{B(\delta ,a)J_1(j_{1,1}B(\delta ,a)r)-L(\delta ,a)J_1(j_{1,1}L(\delta ,a)r)}{r}, 
\end{equation}
and 
\begin{eqnarray}
g(\delta,a;r )|_{r=a}&=&0 \\
\frac{d}{dr}g(\delta,a;r )|_{r=a}&=&2\pi d_{max} \frac{-B(\delta ,a )^2 J_2 (j_{1,1}B(\delta ,a )a)+L(\delta ,a )^2 J_2 (j_{1,1}L(\delta ,a )a)}{r} \\
\frac{d^2}{dr^2}g(\delta,a;r )|_{r=a}&=&6\pi d_{max} \frac{B(\delta ,a )^2 J_2 (j_{1,1}B(\delta ,a )a)-L(\delta ,a )^2 J_2 (j_{1,1}L(\delta ,a )a)}{r^2} .
\end{eqnarray}
Based on the above preparation, we construct the new ASF as follows: 
\begin{equation}
h(a;r)=W_0(a)g(0,a;r)+W_1(a)g(1,a;r),
\end{equation}
where
\begin{equation}
W_0(a )= \frac{|J_2(j_{1,1}a B(0,a )) B(0,a )^2-J_2(j_{1,1}a L(0,a ))L(0,a )^2| }{\sum_{\delta =0}^1|J_2(j_{1,1}a B(\delta ,a )) B(\delta ,a )^2-J_2(j_{1,1}a L(\delta ,a ))L(\delta ,a )^2|},
\end{equation}
and
\begin{equation}
W_1(a )= \frac{|J_2(j_{1,1}a B(1,a )) B(1,a )^2-J_2(j_{1,1}a L(1,a ))L(1,a )^2| }{\sum_{\delta =0}^1|J_2(j_{1,1}a B(\delta ,a )) B(\delta ,a )^2-J_2(j_{1,1}a L(\delta ,a ))L(\delta ,a )^2|} .
\end{equation}
This ASF is designed such that its zeroth-, first-, and second-order differential coefficients at $r=a$ are zero and its energy transmittance is as large as possible.
Therefore, the required pupil function is derived as follows:
\begin{equation}
Q(a;\alpha )=W_0(a)P(0,a;\alpha ) + W_1(a)P(1,a;\alpha ).
\end{equation}